\documentclass[aoas]{imsart}

\RequirePackage{amsthm,amsmath,amsfonts,amssymb}
\RequirePackage[authoryear]{natbib}
\RequirePackage[colorlinks,citecolor=blue,urlcolor=blue]{hyperref}
\RequirePackage{graphicx}
\RequirePackage{bm}
\RequirePackage{url}
\RequirePackage{booktabs}
\RequirePackage{multirow}

\startlocaldefs
\numberwithin{equation}{section}
\endlocaldefs

\begin{document}

\begin{frontmatter}

\title{A Bayesian Spatiotemporal Model to Estimate Disease Burden Using Hospital-Based Active Surveillance}
\runtitle{Estimating Disease Burden from Hospital-Based Active Surveillance}

\begin{aug}
\author{\fnms{Brent}~\snm{Strong}}
\author{\fnms{Claudia}~\snm{Mu\~noz-Zanzi}}
\author{\fnms{Caitlin}~\snm{Ward}}
\end{aug}

\begin{abstract}
Passive surveillance systems, in which data routinely collected by medical facilities are used to monitor the caseload of infectious diseases, are relatively straightforward to implement but often result in underestimation of the burden of disease due to under-diagnosis and imperfect testing. Targeted active surveillance can be used to correct these case counts to better reflect the true burden of disease. However, when the active surveillance effort is performed at a subset of hospitals and passive surveillance data is reported at an aggregated regional level, the resulting spatial misalignment must be reconciled to estimate the true rate of hospital-presenting disease at the spatial region level. Motivated by a recent active surveillance project for leptospirosis in four Puerto Rican hospitals, we address this challenge and develop a novel Bayesian spatio-temporal framework to better reflect the true number of hospital-presenting individuals with the disease. In particular, our method extends the Poisson-logistic framework to incorporate spatial heterogeneity in the probability of  presenting to the hospitals across the study region. Our framework also accounts for imperfect diagnostic testing within the active surveillance data, addressing a common challenge for infectious diseases, particularly for neglected ones like leptospirosis. The model is assessed via simulation under various scenarios and then applied to the motivating leptospirosis data. Our approach offers a comprehensive framework for integrating spatially misaligned passive and active surveillance data, enabling better estimation of true disease burden.
\end{abstract}


\end{frontmatter}

\section{Introduction}

In the context of infectious diseases, it is imperative to accurately quantify the true disease burden, e.g., incidence or mortality rates, across space and time to ensure resources are appropriately deployed to minimize transmission and prevent the worst disease outcomes. However, true counts of disease incidence are virtually never observed due to under-capture by public health surveillance systems. Under-capture is often especially severe for passive surveillance systems, which are the source of most data \citep{murray2016infectious}. This is because these systems require that a patient is seen at a medical facility, the appropriate diagnostic test is ordered, and the test does not produce a false negative result. To better understand disease burden, passive surveillance data may be supplemented with targeted active surveillance studies where individuals are proactively tested. Active surveillance studies may be community-based, where people with the disease are sought out in cities and towns where people live. Alternatively, they may be restricted to selected sentinel hospitals, which is the setting considered in this work \citep{murray2016infectious}.

A common statistical approach to analyzing under-captured case counts is the Poisson-logistic model \citep{WinkelmannZimmermann1993}. In this model, the true, unobserved disease counts are assumed to arise from a Poisson generating process and then cases are captured according to a binomial distribution. The capture probability, which is often modeled using a logistic regression, provides the average fraction of diseased individuals that are counted as cases. When only the under-captured case counts are observed the Poisson mean parameter and capture probability are not separately identifiable, as high incidence rates with low capture probability could produce the same observed counts as low incidence rates with high capture probability. To overcome this challenge, the Poisson-logistic model is often implemented in the Bayesian framework with strongly informative priors to ensure model identifiability \citep{stoner2019hierarchical, bradshaw2024bayesian}. To relax the need for strong priors one can induce identifiability through stricter assumptions on the data generating process (e.g., \cite{lopes2022bias}) or by supplementing the model with a limited number of additional observations in which the true counts are observed (e.g., \cite{dvorzak2016sparse}).

Supplementing the existing passive surveillance system with an active surveillance study is one way to resolve non-identifiability in the context of infectious diseases. If the active surveillance study records whether enrolled subjects were captured by the passive surveillance system, a capture-recapture approach \citep{otis1978statistical} can be used. In the simplest capture-recapture set-up, the two surveillance systems operate independently and simultaneously. Observing the number of cases captured by both systems and only one system permits inference on case capture probabilities and on the parameters of the Poisson disease count generating process. Recent work motivated by pulmonary tuberculosis in Sichuan, China applied this framework to situations in which a community-based active surveillance system only operates alongside passive surveillance at a subset of sites of interest, but disease burden at all sites is of interest \citep{li2020spatial}. \cite{zhang2023hierarchical} generalized their approach to allow modeling of individual level heterogeneity in capture probability. However, these methods do not readily apply to the setting of hospital-based active surveillance encountered by our motivating application, because they assume spatial alignment between the Poisson count generating process and the under-capture process.

In our application of interest, \cite{munoz2025diagnosis} conducted active sentinel surveillance of leptospirosis in four Puerto Rican hospitals and obtained data on patients captured by the independently operating passive surveillance system at those hospitals during the study period. The only other data available were aggregated case counts at the health region level from the passive surveillance system. The fact that we do not observe hospital-level passively surveilled case counts for facilities that were not part of the active surveillance study represents a significant spatial misalignment problem if we suspect that the probability of case capture by the passive surveillance system differs across hospitals.

To illustrate the issues at play, suppose that the probability of case capture by the passive surveillance system is strongly associated with whether a hospital is located in an urban or rural setting. It is easy to envision a scenario in which one region is largely rural and another region is very urban. Thus, we would expect the magnitude of the under-counting to differ substantially between these two health regions because the types of hospitals most patients present to would be different. This calls for methods that can link hospital-level estimates of passive surveillance case capture probabilities with aggregate case counts and account for heterogeneity between regions.

A second limitation of existing models in this setting is that they do not account for imperfect diagnostic testing, a pervasive issue in the setting of neglected infectious diseases like leptospirosis \citep{valente2024diagnosis}. Specifically, diagnostics for these diseases often exhibit low sensitivity, resulting in a high rate of false negatives among truly infected individuals. Consequently, when integrating active surveillance data, failing to incorporate this diagnostic uncertainty and assuming that only those testing positive are diseased may underestimate the true disease burden. We address this limitation using the approach of \cite{ward2023individual} to incorporate individual disease status indicators as parameters in our model. We inform their estimation with strong priors on the sensitivity and specificity of the diagnostic tests. An important advantage of our approach over that of \cite{ward2023individual} is that we are able to marginalize out these indicators, producing a valid model that is more computationally efficient. 

Addressing the two major methodological challenges presented by our application allows for estimation of the parameters of a spatiotemporal regression model for the true disease rate. This regression permits quantification of the true region-level burden of disease and investigation of factors associated with it. In particular, our estimand of interest is the burden of hospital-presenting disease, since both surveillance systems only operate at the hospital level. While community-wide incidence captures total burden including mild or asymptomatic infections, hospital-presentation offers a highly relevant indicator for public health decision-making. By focusing on individuals with symptoms severe enough to require medical intervention, we quantify the population driving healthcare utilization and clinical resource demand. We generally express burden using the disease rate, i.e., the number of hospital-presenting cases per 100,000 people in the general population. 

The remainder of the paper outlining our work proceeds as follows. In Section 2, we discuss our motivating data application in further detail. In Section 3, we outline our Bayesian modeling framework. In Section 4, we describe a simulation study to provide insight into model performance. In Section 5, we conduct the real data analysis. Finally, in Section 6, we provide a discussion and concluding remarks. 
All code for this work will be publicly available at the following URL: \url{https://github.com/brent-strong/lepto_burden}.

\section{Motivating Data} \label{sec:leptodata}

\begin{figure}[!htb] 
    \centering
    \includegraphics[width=0.8\linewidth]{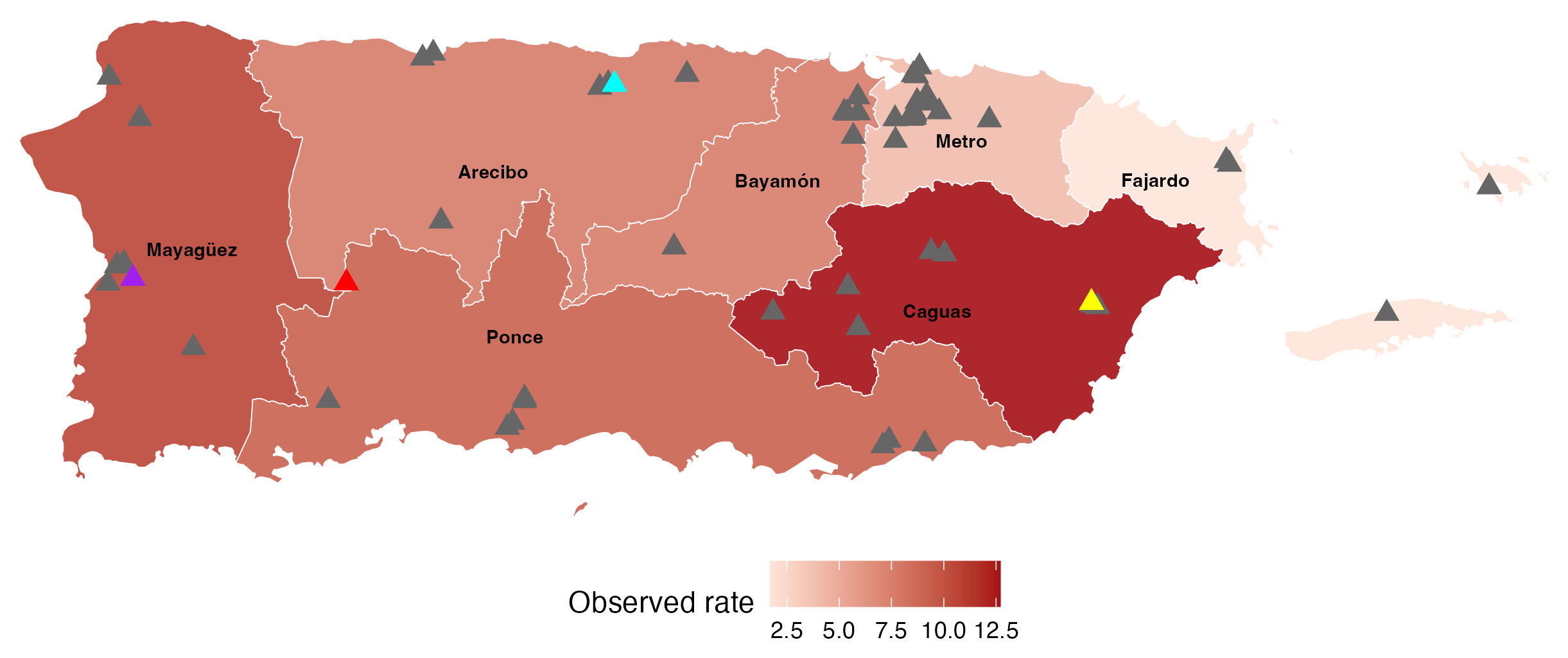}
    \caption{The seven health regions of Puerto Rico with all hospitals denoted by triangles. Hospitals not included in the active surveillance study are shown in gray. Hospitals in the active surveillance study are shown in color. Moving from left to right, the purple triangle denotes Hospital Bella Vista, the red triangle denotes Casta\~ner General Hospital, the cyan triangle denotes Doctor's Center Hospital, and the yellow triangle denotes Ryder Memorial Hospital. Coloring corresponds to the average observed rate of leptospirosis for 2022 and 2023 per 100,000 people.}
    \label{fig:four_hosp}
\end{figure}

Leptospirosis is an environmentally transmitted infectious disease caused by pathogenic strains of the bacterial genus \textit{Leptospira} \citep{bradley2023leptospirosis}. Current estimates indicate that leptospirosis is responsible for approximately 60,000 annual deaths worldwide with low income tropical regions bearing the majority of the disease burden \citep{costa2015global}. Most cases in the United States occur in Puerto Rico \citep{adams2017summary} and a prior study estimated an annual incidence rate of 2.11 per 100,000 people \citep{ramos2018analisis}. However, due to poor case capture, this figure likely represents a substantial under-estimate of the true disease burden in Puerto Rico. Identifying leptospirosis cases is especially challenging due to the non-specificity of clinical presentation (often manifesting as undifferentiated acute febrile illness (UAFI)), inadequate clinician awareness, sophisticated diagnostic laboratory resource requirements, and the imperfect nature of laboratory testing \citep{bradley2023leptospirosis}. The existing passive surveillance system in Puerto Rico relies on the mandatory reporting of all laboratory-diagnosed cases by health facilities. To better understand disease burden, \cite{munoz2025diagnosis} conducted sentinel active surveillance in four Puerto Rican emergency departments, the locations of which are shown in Figure \ref{fig:four_hosp} with boundaries denoting the seven health regions of Puerto Rico. The hospitals included in the active surveillance study were intentionally chosen to be spatially dispersed to account for heterogeneity in capture probabilities and leptospirosis rates.

Starting in July 2019 and ending in May 2021, study recruiters approached patients meeting the criteria of UAFI in each of the four emergency departments. Those that consented were tested using a standardized diagnostic algorithm that combined polymerase chain reaction (PCR) testing and serum antibody testing (IgM and microagglutination testing). Investigators also received reports of any patients that were identified independently by the passive surveillance system during the active surveillance period. The data generated by the active surveillance study is shown in Table \ref{tab:as_table}. In total, 406 UAFI patients were enrolled and tested with one or two diagnostics. 16 patients had at least one positive test. Of those with a positive test, one was also captured by passive surveillance. There were 13 additional patients captured by passive surveillance during the overall period of active surveillance. These patients were identified through referral by the diagnostic laboratory. Also available are publicly reported case counts for 2022 and 2023 at the health region level, which are mapped in Figure \ref{fig:four_hosp} and provided in Table S3.1 of the Supplement. 

\begin{table}[h!]
\centering
\begin{tabular}{lllccccc}
\toprule
\multirow{2}{*}{\textbf{Hospital}} & \multirow{2}{*}{\begin{tabular}[c]{@{}l@{}}\textbf{Health} \\ \textbf{Region}\end{tabular}} & \multirow{2}{*}{\begin{tabular}[c]{@{}l@{}}\textbf{Number} \\ \textbf{Tested}\end{tabular}} & \multicolumn{3}{c}{\textbf{Positive test results}} & \multirow{2}{*}{\begin{tabular}[c]{@{}c@{}}\textbf{Passive \&} \\ \textbf{Active}\end{tabular}} & \multirow{2}{*}{\textbf{Passive}} \\
\cmidrule(lr){4-6}
 & & & \textbf{IgM} & \textbf{PCR} & \textbf{Both} & & \\
\midrule

\multirow{2}{*}{\begin{tabular}[c]{@{}l@{}}Hospital Bella Vista\\ \footnotesize Feb. 2020--Mar. 2020\end{tabular}} & Arecibo & 1 & 0 & 0 & 0 & 0 & 0 \\
 & Mayag\"uez & 26 & 0 & 0 & 0 & 0 & 0 \\
\midrule

\multirow{4}{*}{\begin{tabular}[c]{@{}l@{}}Casta\~ner General Hospital\\ \footnotesize Dec. 2019--Mar. 2020;\\ \footnotesize Jun. 2020--Apr. 2021\end{tabular}} & Arecibo & 20 & 0 & 1 & 0 & 0 & 1 \\
 & Mayag\"uez & 4 & 0 & 0 & 1 & 0 & 0 \\
 & Metro & 1 & 0 & 0 & 0 & 0 & 0 \\
 & Ponce & 48 & 1 & 1 & 0 & 0 & 1 \\
\midrule

\multirow{5}{*}{\begin{tabular}[c]{@{}l@{}}Doctor's Center Hospital\\ \footnotesize Jul. 2019--Mar. 2020;\\ \footnotesize Jun. 2020--Mar. 2021\end{tabular}} & Arecibo & 154 & 2 & 0 & 0 & 1 & 3 \\
 & Bayam\'on & 13 & 0 & 1 & 0 & 0 & 0 \\
 & Fajardo & 1 & 0 & 0 & 0 & 0 & 0 \\
 & Mayag\"uez & 2 & 0 & 0 & 0 & 0 & 0 \\
 & Ponce & 1 & 0 & 0 & 0 & 0 & 0 \\
\midrule

\multirow{3}{*}{\begin{tabular}[c]{@{}l@{}}Ryder Memorial Hospital\\ \footnotesize Mar. 2020; \\ Jun. 2020--May 2021\end{tabular}} & Bayam\'on & 1 & 0 & 0 & 0 & 0 & 0 \\
 & Caguas & 134 & 5 & 1 & 3 & 0 & 8 \\ \\
\bottomrule
\end{tabular}
\caption{Active surveillance data from four hospitals. ``Tested" refers to the number of people tested at the given hospital from the specified health region. Hospital/health region combinations with the number of people tested equal to zero are not shown. The following three columns provide the number testing positive for only the rapid IgM test, only the PCR test, or both. Some that tested positive for only one were not administered the other test. ``Passive \& Active" refers to the number of leptospirosis cases captured by passive surveillance and captured by active surveillance. ``Passive" is total number of leptospirosis patients captured by passive surveillance. Active surveillance was interrupted from March 2020 to June 2020 because of the COVID-19 pandemic, where applicable.}
\label{tab:as_table}
\end{table}

The motivating data exemplifies the two major methodological challenges we seek to address. Both active and passive surveillance operated at the hospital level.  However, the only  external data are aggregate case counts at the health region level. The need for a mechanism that links hospital-level capture probabilities with aggregate case counts becomes apparent when examining Figure \ref{fig:four_hosp}. Estimates of the capture probability that are generated from data for Ryder Memorial Hospital on the east side of the island in Caguas should not be directly applied to correct the reported case counts in Mayag\"uez on the opposite side of the island. As the patients enrolled in the active surveillance study at Ryder Memorial Hospital were primarily from the Caguas region, the estimated capture probability at that hospital should have a larger impact on our understanding of the under-capture process in Caguas. 

Moreover, it is important to account for imperfect sensitivity and specificity for leptospirosis diagnostics. In one study, \cite{niloofa2015diagnosis} found that an IgM-based rapid test had 85\% sensitivity and 85\% specificity for acute leptospirosis. This implies that we would expect 15\% of true leptospirosis patients to be missed and 15\% of non-leptospirosis UAFI patients to incorrectly test positive in the study. Our approach allows for the appropriate incorporation of uncertainty in this regard. 

\section{Methods}

\subsection{Model Specification}

\subsubsection{The Poisson-Logistic Model}

We begin by briefly introducing the Poisson-logistic framework originally proposed by \cite{WinkelmannZimmermann1993} that serves as the basis for our work. The true disease counts for spatial unit $s$ at discrete time $t$ are denoted by $z_{s,t}$ and assumed to be Poisson distributed with mean parameter $\mu_{s,t}$. Due to under-capture, we do not observe the true counts. Instead, we observe $y_{s,t}$, the under-captured case counts, which are assumed to be binomial realizations from the unobserved true disease counts with associated capture probability $\pi_{s,t}$. Therefore, the Poisson-logistic model can be written as

\begin{equation}
\label{eq:l_pogit}
\begin{split}
z_{s,t} &\sim \text{Poisson}(\mu_{s,t}), \quad \log(\mu_{s,t}) = \beta_0 + \sum_{j=1}^J \beta_j x_{j,s,t}\\
y_{s,t} \mid z_{s,t} &\sim \text{Binomial}(z_{s,t}, \pi_{s,t}), \quad \text{logit}(\pi_{s,t}) = \alpha_0 + \sum_{k=1}^K \alpha_k w_{k,s,t} 
\end{split}
\end{equation}

\noindent where $\{x_{s,t}\}$ is a set of covariates that affect the trajectory of the true disease counts and $\{w_{s,t}\}$ are a set of covariates that are relevant to the under-capture process. Both regressions are easily extended to incorporate spatial or temporal effects as in \cite{stoner2019hierarchical} as well as appropriate offsets. Under this model, it is straightforward to show that after marginalizing over $z_{s,t}$, the distribution of $y_{s,t}$ is Poisson distributed with mean parameter $\mu_{s,t}\times\pi_{s,t}$. Estimation of $\mu_{s,t}$ in this framework permits inference on functions of $z_{s,t}$ like rates, our estimand of interest.

\subsubsection{Adapting the Poisson-Logistic Model}

Direct application of the Poisson-logistic framework to the under-captured counts in the motivating application is not straightforward. The primary difficulty is that active surveillance capture occurs at the hospital level, while $y_{s,t}$ is aggregated at the spatial unit level. This introduces a spatial misalignment problem \citep{banerjee2003hierarchical}, as patients from a single spatial unit can typically present to many different hospitals, which may or may not be located in the same spatial unit where they reside. Hospitals may differ in their capture probabilities, so there is no single probability parameter $\pi_{s,t}$ governing the under-capture process for each spatial unit and time point. Instead, we decompose the capture probability into the probability of capture conditional on presenting at a given hospital, multiplied by the probability of presenting at that hospital. Thus,  we use the following distribution for $y_{s,t}$:
\begin{align}
y_{s,t} &\sim \text{Poisson}\left(\mu_{s,t} \times \sum_{h=1}^H \pi_{h, s}^{C}\pi_{h}^{P}\right), \label{eq:disty}
\end{align}
where $\pi_{h}^{P}$ is the probability that a disease case presenting to hospital $h$ (among $H$ total hospitals) is captured by the passive surveillance system and $\pi_{h,s}^{C}$ is the average probability of an individual from spatial unit $s$ choosing to present at hospital $h$. The consequence of this specification is that the mean parameter is multiplied by a weighted sum of the passive surveillance case capture probabilities for each of the hospitals. If there was only one hospital a patient from a given spatial unit could go to, this model would reduce to the typical Poisson-logistic setup shown in Equation \ref{eq:l_pogit}. A more formal justification for this distribution for $y_{s,t}$ is described in Section S1.1 of the Supplement.

\subsubsection{Hospital Choice Probabilities}

\begin{figure}[!htb] 
    \centering
    \includegraphics[width=0.8\linewidth]{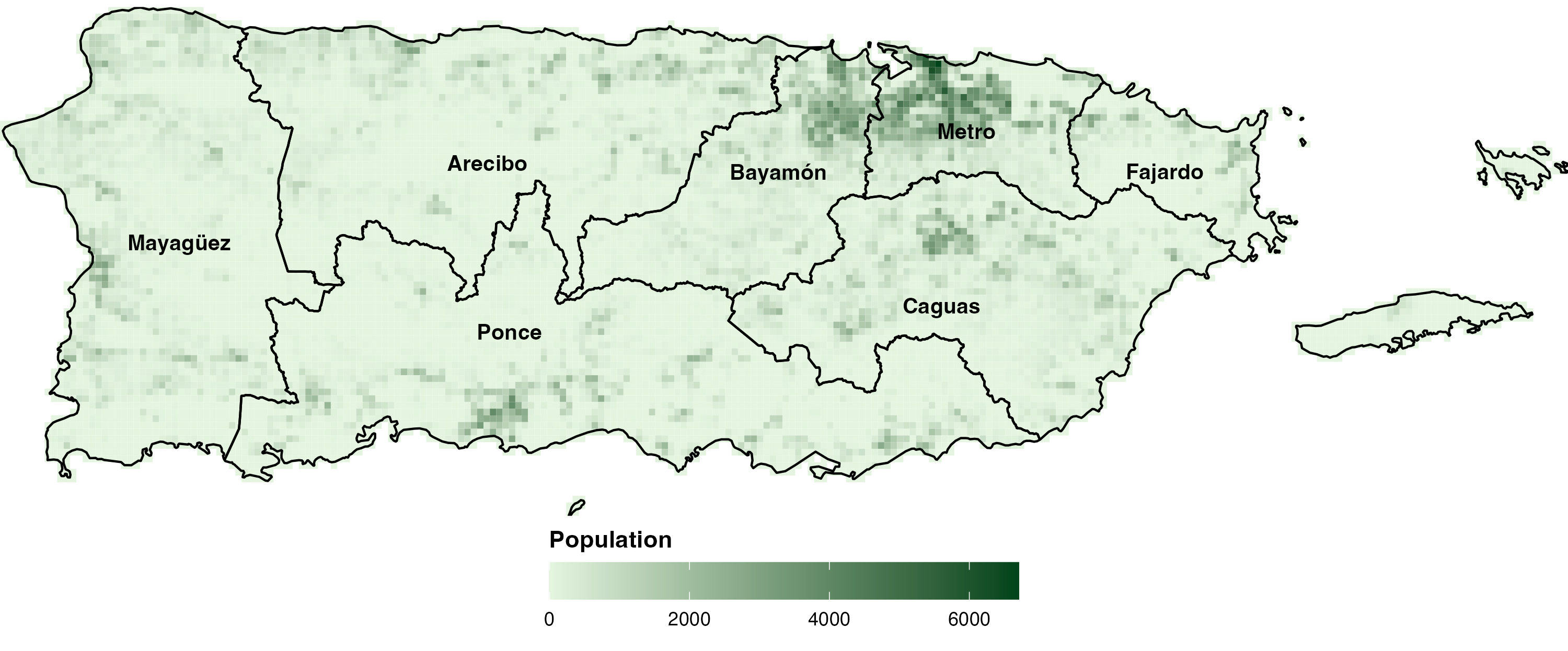}
    \caption{The seven health regions of Puerto Rico with the one kilometer spatial subunits used in the simulation study and data analysis. Spatial subunits are colored darker green if they have a higher population (higher $N_{u_s}$). The most populous area corresponds to San Juan.}
    \label{fig:spat_sub}
\end{figure}

In order to define the probability of choosing hospital $h$ in spatial unit $s$ ($\pi_{h,s}^C$) we partition each spatial unit $s$ into $U_s$ small subunits. The purpose of this partitioning is to account for spatial heterogeneity in hospital choice \textit{within} each spatial unit. The actual subunits we use are shown in Figure \ref{fig:spat_sub}. 

\noindent Using these partitions, we assume that the probability of choosing hospital $h$ ($\pi_{u_s}^{C,h}$) for a disease case is constant in each subunit and is of the following form:

\begin{align}
\pi_{h,u_s}^{C} = \frac{\text{capacity}_h^{\delta}/\text{distance}_{h,u_s}^{\gamma}}{\sum_{l=1}^{H}\text{capacity}_l^{\delta}/\text{distance}_{l,u_s}^{\gamma}}. \label{eq:gravity}
\end{align}

\noindent In this equation, $\text{distance}_{h,u_s}$ is a measure of distance from spatial subunit $u_s$ to hospital $h$, and $\text{capacity}_h$ is some measure of the capacity of hospital $h$ (e.g., number of beds). Higher values of $\gamma$ imply that the probability of choosing hospital $h$ decreases faster as distance increases; higher values of $\delta$ imply that the probability of choosing hospital $h$ increases faster as capacity increases. The specified form of choice probabilities resembles a gravity model, which is often used to model accessibility to healthcare facilities \citep{joseph1982measuring}. A formal derivation of the hospital choice probability can be provided by applying discrete choice methods from economics \citep{train_discrete_2009}, and this is provided in Section S1.2 of the Supplement. Combining across spatial subunits, the weight on hospital $h$ in Equation \ref{eq:disty} is:

\begin{align}
\pi_{h,s}^C =N_s^{-1}\sum_{u=1}^{U_s} N_{u_s} \pi_{h,u_s}^{C} \label{eq:avg_c_prob},
\end{align}

\noindent where $N_s$ is the population of spatial unit $s$, and $N_{u_s}$ is the population in spatial subunit $u_s$. 

In our analysis, we use the set of one kilometer by one kilometer spatial subunits constructed by \cite{pfeffer_malariaatlas_2018} and \cite{weiss_global_2020} as part of the Malaria Atlas project, which provides worldwide friction surfaces for the purpose of calculating motorized or ambulatory travel times to healthcare facilities. These surfaces have been previously used to predict hospital catchment areas \citep{millar2025estimating}. We use the friction surfaces to quantify distance in terms of motorized travel time from the subunit centroid to the hospital since potential patients are more likely to think in these terms than in Euclidean distance. To identify hospitals and their capacity, we use the set of hospitals in Puerto Rico compiled by the Department of Homeland Security \citep{dhs2019hospitals}. We estimate the population of each subunit by using the EPA's EnviroAtlas dasymetric population map \citep{baynes2022improving}. 

\begin{figure}[!htb] 
    \centering
    \includegraphics[width=0.8\linewidth]{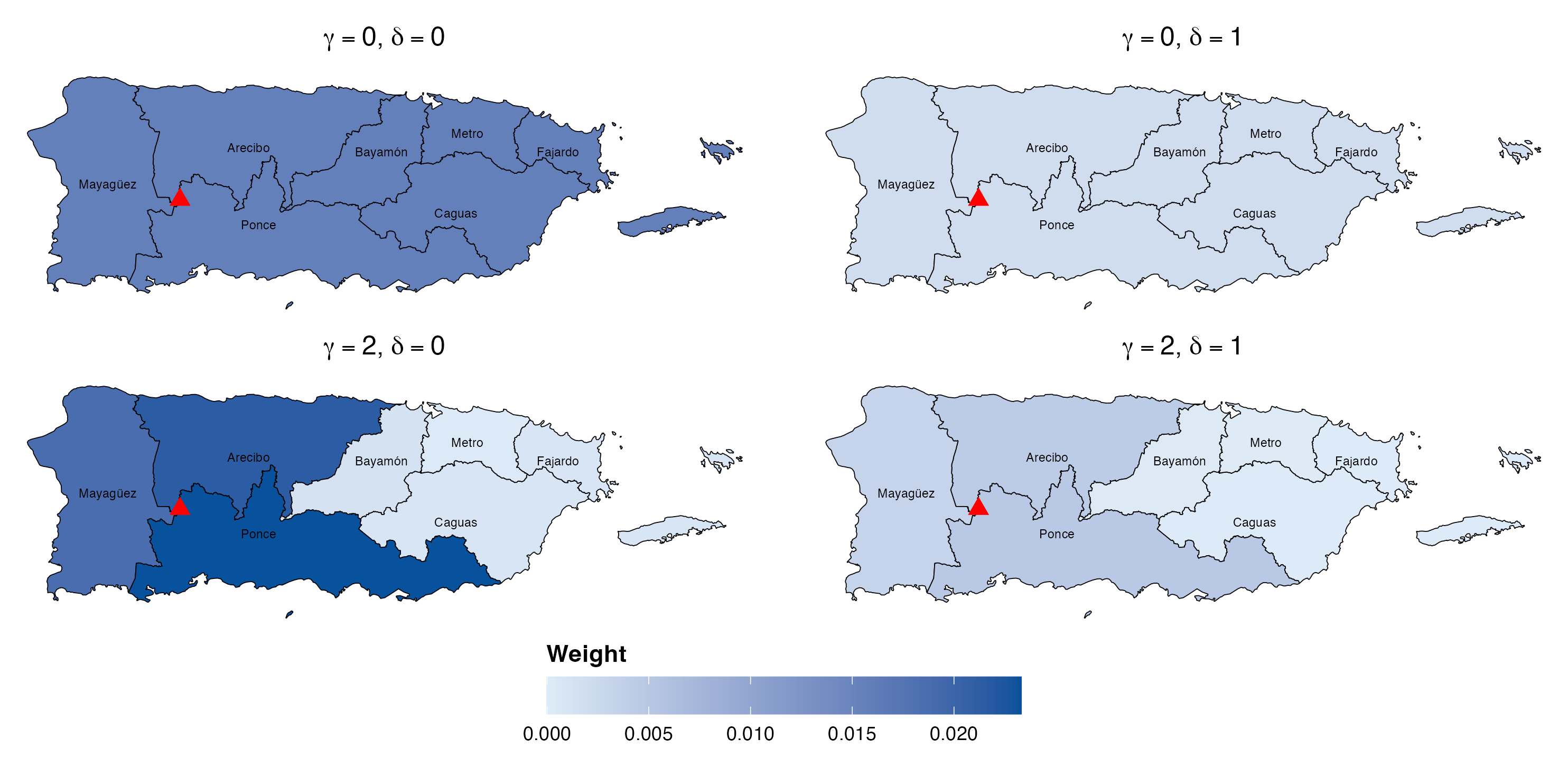}
    \caption{Average probability of choosing to present ($\pi_{h,s}^C $) at Casta\~ner General Hospital for each health region. The top left panel corresponds to a situation in which all hospitals are weighted equally in each region's adjustment factor. In the top right panel, only capacity is considered. In the bottom left, only distance is considered. Finally, in the bottom right both distance and capacity are factored in.}
    \label{fig:gravity}
\end{figure}

Figure \ref{fig:gravity} shows the estimated average probability ($\pi_{h,s}^C$) of a patient from spatial unit $s$ choosing to present at Casta\~ner General Hospital for different values of $\gamma$ and $\delta$. If we set $\gamma=0$ and $\delta=0$, then $\pi_{h,s}^{C} = 1/H$ for all spatial units, which is equivalent to not using a gravity model at all. This is plainly problematic, as it would imply that the average probability of choosing Casta\~ner General Hospital is the same in  Mayag{\"u}ez and Fajardo, even though Fajardo is on the complete opposite side of the island. Another interesting case is if we set $\gamma=0$ and $\delta=1$. Then the average probability of choosing hospital $h$ is equal to it's proportion of total hospital capacity. For Casta\~ner General Hospital, including capacity results in large reductions in the associated choice probability since it has only 24 beds. Although this is an improvement, the average choice probability for this hospital remains constant across all spatial units. Thus, it is important to factor in distance as well. The bottom two panels of Figure \ref{fig:gravity}, which factor in distance by setting $\gamma = 2$, show higher average probabilities of choosing Casta\~ner General Hospital for nearby spatial units. As before, the probability of choosing this specific hospital is reduced when capacity is incorporated and $\delta = 1$.

Because estimation of $\gamma$ and $\delta$ is difficult in our setting (see Section S1.2 of the Supplement for further discussion), we recommend fixing them at reasonable values or using external data sources to estimate them.  We chose to use $\gamma=2$ and $\delta=1$ in the simulation study as that specification has been previously used for modeling the interaction between spatial units in a compartmental model of cholera dynamics \citep{wheeler2024informing}. Because there was a lot of information missing on hospital capacity, we set $\delta=0$ in the real data analysis. We anticipate that the sensitivity of the results to reasonable values for $\gamma$ and $\delta$ will be limited.

\subsection{Addressing Identifiability with Active Surveillance Data} \label{sec:asd}

As previously discussed, the central difficulty in the Poisson-logistic framework is the non-identifiability of $\mu_{s,t}$ and the passive surveillance capture probabilities, $\pi^P_{h}$. We formulate an approach similar to \cite{li2020spatial} that allows us to incorporate data from multiple surveillance systems, but note that our Bayesian approach to inference allows for informative prior distributions when necessary. We assume that active and passive surveillance systems operate independently at the hospital-level, and consider data in unique bins defined by hospital, spatial unit, and time. Although we have dropped these indexes for the remainder of this section, they are implicit.

For each hospital in the active surveillance study, the observed data provide $n^P$, the number of disease cases captured by passive surveillance (corresponding to the ``Passive" column in Table \ref{tab:as_table}), $n^{A}$, the number of disease cases captured by active surveillance, and $n^{PA}$, the number of disease cases captured by both passive and active surveillance (corresponding to the ``Passive \& Active" column in Table \ref{tab:as_table}). Let $n^D$ denote the true number of disease cases, which is never observed due to imperfect capture by any realistic surveillance system. From these quantities, we can construct four counts that enumerate all possibilities of capture across both surveillance systems: captured by both systems ($n^{PA}$), captured by only the passive system ($n^P-n^{PA}$), captured by only the active system ($n^{A}-n^{PA}$), or captured by neither system ($n^D - n^P - n^{A} + n^{PA}$). Concatenating the outcomes and the associated probabilities into the vectors $\mathbf{n} = \{n^{PA}, n^P-n^{PA}, n^{A}-n^{PA}, n^D - n^P - n^{A} + n^{PA}\}$ and $\boldsymbol{\pi} = \{\pi^{P}\times \pi^{A}, \pi^{P}\times (1-\pi^{A}), (1-\pi^{P})\times \pi^{A}, (1-\pi^{P})\times (1-\pi^{A})\}$, respectively, the capture-recapture framework defines a multinomial likelihood for the observed hospital-level data, written as 

\begin{align}
\mathbf{n} \sim \text{Multinomial}(n^D,\boldsymbol{\pi}). \label{eq:crc}
\end{align}

\noindent This multinomial likelihood permits inference on the three unknowns: $n^D$, $\pi^{P}$, and $\pi^{A}$. We will generally include an index for hospital for $\boldsymbol{\pi}$ since passive surveillance capture probability is assumed to differ across hospitals.

\subsection{Incorporating Diagnostic Test Uncertainty in Active Surveillance Data}

Previous work has generally assumed that the number of diseased individuals captured by active surveillance, $n^{A}$, is observed even though it virtually never is because infectious disease diagnostic tests are imperfect. Estimating $n^{A}$ from the multinomial likelihood alone is problematic because it introduces a fourth unknown, leading to identifiability issues. Fortunately, test results provides information on the value of $n^{A}$.  We denote the $n^T \times K$ matrix of test results as $\mathbf{Y}$ where $Y^{j,k}\in \{0, 1\}$ indicates a negative or positive result for test $k = 1, ..., K$ given to patient $j = 1, ..., n^T$. $K$ is the number of unique tests given as part of active surveillance and $n^T$ is the number of patients tested under active surveillance (corresponding to the ``Tested" column in Table \ref{tab:as_table}). This number will generally be much larger than the true number of individuals with the disease that are captured by active surveillance ($n^{A}$). We denote the vector of sensitivities for all tests by $\boldsymbol{\nu}$ and the vector of specificities by $\boldsymbol{\kappa}$. Assuming that tests are conditionally independent given the disease status of all tested individuals $\mathbf{d} =\{d_1, ..., d_{n^T}\}$, the probability of the complete set of test results can be written as $\Pr(\mathbf{Y}|\boldsymbol{\nu},\boldsymbol{\kappa},\mathbf{d}) = \prod_{j=1}^{n^T}\prod_{k = 1}^K \Pr(Y_{jk}|\nu_k,\kappa_k,d_j)$. This probability is a function of the sensitivity and specificity of each test, depending on the underlying disease status of each individual. The full form is provided in Section S1.3 of the Supplement and takes the same form as that used by \cite{ward2023individual}. In subsequent sections, we show how the testing data likelihood permits inference on $n^{A}$.

In truth, $n^P$ is not perfectly observed either since tests do not have perfect specificity (i.e., they may falsely identify some patients as a disease case). Because accounting for a passive surveillance system that includes some false positives would require additional data on passive surveillance testing patterns, we generally assume that all cases captured by passive surveillance are correctly classified. This is a defensible assumption since capture by the passive surveillance system generally requires both clinician suspicion of infection and a positive test.

\subsection{Bayesian Model Specification} \label{sec:model}

We have now described all of the building blocks necessary to formulate a Bayesian model that will allow us to estimate the true disease rates. We describe the joint distribution for the data, latent quantities, and model parameters, which is equivalent to the posterior up to a normalizing constant. By formulating a valid joint distribution, we can then sample all unknown quantities from the posterior using Markov chain Mone Carlo (MCMC) methods. We denote the vector of all parameters except $n^{A}$ and $n^D$ by $\Theta$, which includes all parameters describing the Poisson mean, $\mu_{s,t}$(e.g., $\boldsymbol{\beta}$), the parameters associated with the hospital-specific probabilities of case capture, $\pi^P_{h}$, and test sensitivity and specificity ($\boldsymbol{\nu}$ and $\boldsymbol{\kappa}$). The probabilities of hospital choice, $\pi^C_{h, s}$, are not included as they are fixed. Latent quantities $n^{A}$ and $n^D$ are denoted by $\mathcal{L}$. Data generated for each unique combination of spatial unit, time, and hospital are considered separately, so we now introduce subscripts to that effect. Let $\mathcal{H}^A_t$ denote the set of hospitals under active surveillance at time $t$ and $\mathcal{D}$ the observed data comprised of disease counts and test results.  We write the joint distribution as:

\begin{equation}
\begin{aligned}
\Pr\left(\mathcal{D}, \mathcal{L}, \Theta\right)
&= \prod_{s=1}^S \prod_{t=1}^T
\underbrace{
\Pr \left(y_{s,t}|\{n^P_{h,s,t}\}_{h \in \mathcal{H}^A_t},\Theta\right)
}_{\text{model for aggregate cases}} \\
&\quad \times \prod_{h \in \mathcal{H}^A_t}
\underbrace{
\Pr \left(\mathbf{n}_{h,s,t},\mathbf{Y}_h|\Theta,n^D_{h,s,t} \right)
\Pr\left(n^D_{h,s,t}|\Theta\right)
}_{\text{model for active surveillance data}}
p(\Theta)
\end{aligned}
\label{eq:joint}
\end{equation}

We first consider the model for aggregate cases. Since $y_{s,t}$ is simply the sum of $n^P_{h,s,t}$ across all hospitals, the distribution of $y_{s,t}$ conditional on all $n^P_{h,s,t}$'s from actively surveilled hospitals is:

\begin{align}
\Pr \left(y_{s,t}\mid \{n^P_{h,s,t}\}_{h \in \mathcal{H}^A_t},\Theta\right)
&= \Pr\left(\sum_{h \notin \mathcal{H}^A_t} n^P_{h,s,t} \,\middle|\, \Theta\right) \notag \\
\sum_{h \notin \mathcal{H}^A_t} n^P_{h,s,t}
&\sim \operatorname{Poisson}\left(
\mu_{s,t} \times \sum_{h \notin \mathcal{H}^A_t}\pi^C_{h,s}\pi_{h}^{P}
\right).
\label{eq:aggregate}
\end{align}

\noindent Intuitively, we are only considering case counts from non-actively surveilled hospitals in the model for aggregate case counts. Case counts from actively surveilled hospitals are included in the model for active surveillance data (into which the same Poisson generating process indirectly feeds). Same as before, equation block \ref{eq:aggregate} can be justified by the data generating model described in Section S1.1 of the Supplement.

The second piece of the joint distribution to discuss is the model for active surveillance data. We can factorize the distribution for the vector of possible outcomes for capture by active and passive surveillance at each hospital and the test results for those tested by active surveillance as:

\begin{align}
\Pr \left(\mathbf{n}_{h,s,t},\mathbf{Y}_h|\Theta,n^D_{h,s,t} \right) = \underbrace{\Pr(\mathbf{n}_{h,s,t}|\Theta,n^D_{h,s,t})}_{\text{capture-recapture model}} \underbrace{\Pr(\mathbf{Y}_{h,s,t}|\boldsymbol{\nu},\boldsymbol{\kappa},\mathbf{n}_{h,s,t})}_{\text{model for testing data}} 
\end{align}

\noindent The capture-recapture model takes the multinomial form shown in equation \ref{eq:crc}. The model for testing data can be written as
\begin{align}
Pr(\mathbf{Y}_{h,s,t}|\boldsymbol{\nu},\boldsymbol{\kappa},\mathbf{n}_{h,s,t}) = \sum \Pr(\mathbf{Y}_{h,s,t}|\boldsymbol{\nu},\boldsymbol{\kappa},\mathbf{d}_{h,s,t})\Pr(\mathbf{d}_{h,s,t}|\mathbf{n}_{h,s,t}), \label{modelY}
\end{align}
where $\mathbf{d}_{h,s,t}$ is a latent vector of length $n^T_{h,s,t}$ containing the true disease status of each patient and the sum is across $n_{h,s,t}^T-n_{h,s,t}^{PA} \choose n_{h,s,t}^{A} - n_{h,s,t}^{PA}$ possible disease status vectors. For the remainder of this section, we drop the subscripts $s$, $h$, and $t$ to prevent the notation from becoming too dense.

Since we assume that the specificity of the passive surveillance process is $1$, we constrain the first $n^{PA}$ disease status indicators in $\mathbf{d}$ to be $1$ ($n^{PA}$ denotes the number of patients identified by both active and passive surveillance). A na\"ive approach that enumerates all possible combinations is infeasible for large enough values of $n^T$, so we use a recursive algorithm that makes computation tractable. Full details of the algorithm are provided in Section S1.3 of the Supplement. This algorithm permits us to avoid drawing $\mathbf{d}$ explicitly as a parameter and instead marginalize over it. We set $\Pr(\mathbf{d}|\mathbf{n})= {n^T-n^{PA} \choose n^{A} - n^{PA}}^{-1}$ to reflect the assumption in our model that patients tested as part of active surveillance in groups defined by hospital, spatial unit, and time are exchangeable. Thus, conditional on a total number of tested patients with the disease, any combination of patients having the disease consistent with this total number is equally probable prior to seeing test results. 

We generally do not directly use the model for active surveillance. Rather, we follow \cite{royle2004n} and \cite{li2020spatial} and marginalize over $n^D$. We expand this to also marginalize over $n^{A}$. We refer to this as the marginal model for active surveillance which takes the following form:

\begin{align}
\sum_{n^{A}=n^{PA}}^{n^T}\sum_{{n}^D=n_{h}^{A}+n^P-n^{PA}}^{\infty}\Pr(\mathbf{Y}|\boldsymbol{\nu},\boldsymbol{\kappa} ,\mathbf{n})\Pr(\mathbf{n}|\Theta,n^D)\Pr(n^D|\Theta) \label{mmas}
\end{align}

\noindent The advantage of this marginal model is that it makes computation more straightforward by reducing the number of parameters that need to be sampled as part of the MCMC scheme. Although the second sum technically goes to infinity, it can be truncated at a large number. 

The data generating model outlined in Section S1.1 of the Supplement implies the following prior distribution for $n^D$:

\begin{align}
n^D_{h,s,t}|\Theta &\sim \text{Poisson}\left(\mu_{s,t} \times \pi_{h,s}^C\right) \label{eq:nc}
\end{align}

\noindent  This distribution reflects our assumption that if we expect $100$ cases for the entire spatial unit $s$ (as an example) and the average probability of presenting to hospital $h$ is $0.1$, we would expect to see $10$ patients from $s$ with the disease at hospital $h$. 

\subsection{Estimating the True Disease Rate with Posterior Predictive Distributions} \label{sec:ppd}

Our estimand of interest is the true rate of hospital-presenting disease per 100,000 people among the general population for a given spatial unit $s$ at a given time point $t$. It can be computed from estimates of true counts as $z_{s,t}/N_s$ if $N_s$ is measured in units of 100,000 people. Estimating true rates of the disease requires the formulation of posterior predictive distributions. We assume that active surveillance is conducted for times $t \in \{1,..., T\}$. Aggregate case counts to be corrected may or may not be available at these times as well. Mirroring the motivating application, we are interested in the posterior predictive distribution of true rates for time period $T+1$, for which we have only observed the under-captured aggregate counts $y_{s,T+1}$ from the passive surveillance system. Using the relation $z_{s,T+1} = y_{s,T+1} + (z_{s,T+1} - y_{s,T+1})$ , we can draw a sample from the posterior predictive distribution for $z_{s,T+1}$ by drawing a sample of $z_{s,T+1} - y_{s,T+1}$ from the following distribution for each posterior draw of the parameters:

\begin{align}
\label{eq:ppd}
z_{s,T+1} - y_{s,T+1} &\sim \text{Poisson}\left(\mu_{s,T+1} \times \sum_{h=1}^H \left(1-\pi^P_{h}\right) \pi^C_{h,s}\right).
\end{align}
\noindent The derivation is presented in Section S1.1 of the Supplement. After drawing the number of unobserved individuals with the disease from this distribution, calculation of rates is straightforward since $y_{s,T+1}$ is observed. We have also considered posterior predictive distributions for true disease counts in the setting where the aggregated case counts are reported during the time periods in which active surveillance is operating at a subset of hospitals (i.e., $t \in \{1,..., T\}$), but reserve discussion of this scenario for Section S1.4 of the Supplement. 

\subsection{Computation and Prior Distributions}

Because the posterior of our Bayesian model is analytically intractable, MCMC methods are required. We use the Nimble package to implement the model and sample from the posterior distribution \citep{de2017programming}. This required development of a custom likelihood function to marginalize out individual disease indicators as described in Section \ref{sec:model}. We found that default random walk samplers generally worked well and that a non-centered parameterization of latent random effects greatly improved convergence as described by \cite{papaspiliopoulos2007general}. We achieved additional improvements when we used elliptical slice sampling for multivariate Gaussian random effects \citep{murray2010elliptical}. For each model, we ran multiple chains with a 10,000 iteration burn-in period and 40,000 iterations saved for evaluation of posterior distributions. Convergence was assessed by examining trace plots and the $\hat{R}$ statistic.

In terms of priors, it is typically necessary to place informative priors on the sensitivity and specificity parameters to ensure identifiability. We found that beta-distributed priors were a sensible choice. For variance parameters, weakly informative log-normal priors aided convergence. Finally, we use normal priors on intercepts and regression coefficients which can be made appropriately informative by setting the standard deviation hyperparameter to the desired value. 

\section{Simulation Study}

\subsection{Simulation Design} \label{sec:sim_design}

We conducted a simulation study to evaluate the performance of our model under known conditions, choosing  specifications that mimicked the motivating data analysis where possible. The following data generating models were used for all scenarios:

\begin{equation}
\label{eq:spreg}
\begin{split}
\log(\mu_{s,t}) &= \beta_0 + \beta_1 Rain_{s,t} + \Phi_s + \tau_s + \log(N_s) \\
\boldsymbol{\Phi} &\sim \text{ICAR}(\sigma^2_\Phi),\;
\boldsymbol{\tau} \sim \mathcal{N}(0,\sigma_{\tau}^2 I)
\end{split}
\end{equation}

\begin{equation}
\label{eq:probreg}
\begin{split}
\text{logit}(\pi_{h}^{P}) &= \alpha_0 + \phi_{h}, \;
\boldsymbol{\phi} \sim \mathcal{N}(0,\Sigma^\phi) \\
\Sigma^\phi_{h,h'} &= \sigma_\phi^2 \exp(-\rho d_{h,h'}), \;
\rho = 0.06
\end{split}
\end{equation}

\noindent We set $\beta_0$ equal to the mean of the log-transformed observed rates for 2023 after multiplication by a factor of $\frac{1}{\text{logit}^{-1}(\alpha_0)}$ ($\alpha_0$ is the passive surveillance case capture probability for an average hospital). $Rain_{s,t}$ denotes centered cumulative annual average rainfall in spatial unit $s$ at time $t$, calculated from NCEP Stage IV rainfall accumulation data \citep{LinMitchell2005}. Thus, $\beta_1$ is a scalar whose true value we set to correspond to a 10\% increase in the leptospirosis rate for every 5 additional inches of rainfall in that year. Rainfall is known to be related to \textit{Leptospira} pathogen load in the environment \citep{thibeaux2024rainfall}. $\Phi_{s}$ is a spatially correlated  effect whose standard deviation $\sigma_\Phi$ we set to $0.5$. Spatial correlation is enforced through an intrinsic conditional autoregressive (ICAR) prior \citep{rue2005gaussian}. Under this prior, the conditional distribution of each spatial random effect depends on the mean of the random effects in neighboring areas, inducing smoothing across adjacent regions through the specified neighborhood structure. We classify health regions that share a boundary as neighbors. $\theta_{s}$ accounts for residual heterogeneity, and we also set its standard deviation $\sigma_\tau$ to $0.5$. 

The logistic regression model for the passive surveillance capture probability at hospital $h$ consists of an intercept and a zero-mean spatial effect in which $d_{h,h'}$ is the Euclidean distance between hospital $h$ and $h'$. The spatial effect is specified to be multivariate Gaussian with an exponential covariance kernel. $\rho=0.06$ corresponds to an effective range of approximately 50 kilometers (the main island of Puerto Rico is 177 km long). We set $\sigma_\phi=0.5$.

Additional hospital-presenting UAFI patients that could potentially enter the active surveillance study were generated from a Poisson distribution assuming a homogeneous rate of 600 cases per 100,000 people across the entire island. This reflects the fact that leptospirosis patients comprised only 4\% of UAFI patients in the active surveillance data \citep{munoz2025diagnosis}. Thus, we expect the rates of additional UAFI cases to be much higher than that of leptospirosis. Leptospirosis and UAFI patients are assumed to choose hospitals in the same fashion as described by Equation \ref{eq:gravity}, with $\gamma=2$ and $\delta=1$. For the large number of hospitals missing capacity data, we randomly imputed it by sampling from the uniform discrete distribution ranging from the minimum observed capacity to the maximum observed capacity.

We generate data for the scenario in which the four hospitals included in the motivating active surveillance study are surveilled for two years (i.e., $t \in \{2022,2023\}$). We assume that passively captured aggregate counts of leptospirosis cases at the health region level are available for those same years. We vary two factors in the simulation study: the passive surveillance capture probability for an average hospital ($\alpha_0$) and the informativeness of the prior on $\alpha_0$. We examine four values of $\alpha_0$ corresponding to baseline passive surveilance capture probabilities of 10\%, 25\%, 50\%, and 75\%. For each of these values, we consider normal priors centered on the true $\alpha_0$ value with standard deviations on the logit scale equal to $1.5$, $1$, $0.5$, and $0.1$, giving a total of 16 simulation scenarios. This allows us to examine the impact of using increasingly informative prior information for passive surveillance case capture probabilities, which may be important given the small number of hospitals surveilled. For each scenario, we ran 200 simulations. 

In the actively surveilled hospitals, we assume that each leptospirosis patient is captured by active surveillance with 25\% probability ($\pi^{A}=0.25$). This number was chosen because \cite{munoz2025diagnosis} report data that suggests approximately 25\% of patients meeting the pre-screening criteria for study inclusion were ultimately enrolled in the active surveillance effort. Further  discussion of active surveillance enrollment is provided in the Section \ref{sec:disc}. In our simulations, all patients tested as part of active surveillance receive a rapid IgM-based test. They also receive a PCR test with 75\% probability. Diagnostic test accuracy varies by day of illness but under the assumption that most UAFI patients seek care during the first week of illness \citep{munoz2025diagnosis}, we set the IgM-based test to have 85\% sensitivity and 85\% specificity. For the PCR test, we set sensitivity to 80\% and specificity to 95\%. These values were informed by results from the literature \citep{niloofa2015diagnosis, valente2024diagnosis, le2025urine}.

To understand the properties of our model as the active surveillance sample size increases, we consider the four hospital scenario plus three additional hypothetical active surveillance projects in which the existing four hospitals are augmented such that there are at least one, two, and three hospitals under surveillance in each of the seven health regions for a total of 7, 14, and 21 hospitals. For each hypothetical active surveillance study, we randomly and uniformly sample which additional hospitals are put under active surveillance in each health region. We only consider a 25\% passive surveillance capture probability and use a normal prior centered on the true value with a standard deviation of $1$. As before, we ran 200 simulations under each of these additional scenarios. Because sensitivity and specificity are not separately identifiable from capture probabilities (an issue not fully mitigated by the addition of more data), we also varied the informativeness of the priors on sensitivity and specificity parameters to see the impact on model performance. For the sake of simplicity, we set the true value to 0.85 for all sensitivity and specificity parameters. In the base case, we used priors of the form $\text{Beta}(8.5,1.5)$ with a standard deviation of approximately $0.11$. These were the same priors placed on sensitivity and specificity of the IgM-based test in the main simulation study. We then considered priors with the same mean but standard deviations of $0.08$, $0.04$, and $0.01$.

 \subsection{Simulation Evaluation}
 
To evaluate the performance of the model, we sampled an additional identically distributed data set for $t=2024$ but only considered the aggregate, passively captured health region counts. This is meant to mimic the scenario in our motivating data. Under the assumption that there are no time trends in passive surveillance capture probabilities, the rates for 2024 are predicted using the appropriate posterior predictive distributions as described in Section \ref{sec:ppd}.

We evaluated the predicted rates using several performance metrics. We computed the geometric mean percent bias across all seven regions using the median of posterior draws as a point estimate, calculated as $100 \times \left(\exp\left\{\frac{1}{7M}\sum_{m=1}^M\sum_{s=1}^7\log\left(\frac{\tilde{z}_{s,2024}^{m}/N_s}{z_{s,2024}/N_s}\right)\right\}-1\right)$, where $m$ indexes simulations and $\tilde{z}_{s,2024}^{m}/N_s$ is the median posterior prediction for the true rate in spatial unit $s$ in 2024 and $z_{s,2024}/N_s$ is the true rate. We also evaluated the mean coverage probability of 95\% prediction intervals for the true rates across all seven regions. Coverage probability was calculated as the proportion of the 200 simulations where the 95\% credible interval contained the true value, and the mean coverage probability averages these across the seven regions. 

To better measure the performance of the entire posterior predictive distribution, we also calculated the median continuous ranked probability score (CRPS) for island-wide rate predictions and its multivariate generalization, the median energy score (ES), for the spatial unit level rate predictions in 2024. These metrics are commonly used to assess the accuracy of probabilistic forecasts and have the desirable property of being strictly proper, which incentivizes honest forecasts \citep{gneiting2007strictly}. Unlike mean absolute error, these scores evaluate the full predictive distribution rather than only a point prediction, rewarding forecasts that are both well calibrated and sharp. For both CRPS and ES, smaller values indicate better predictive performance, with a value of zero corresponding to a perfect prediction. In Section S2.1 of the Supplement we provide more detail on the calculation of CRPS and ES. In addition to evaluating quality of the posterior predictions, we also evaluated estimation of key parameters by visually examining the posterior means and 95\% credible intervals compared to the true values across simulations. Relevant figures are provided in Section S2.3 of the Supplement. We used the Gelman-Rubin statistic ($\hat{R}$) based on four chains to evaluate convergence \citep{gelman1992inference}. Convergence was defined as all parameters with $\hat{R}<1.1$.

\subsection{Simulation Results}

Convergence rates varied across scenarios. When the standard deviation parameter in the prior for $\alpha_0$ was 0.1, 0.5, or 1, convergence rates were generally high, ranging from 88\% to over 99\%. Unsurprisingly, convergence rates worsened as the prior on $\alpha_0$ became less informative.  Given the low number of surveilled hospitals, the posterior is likely to be highly diffuse with inadequate prior information, making exploration difficult. Factors associated with non-convergence (after adjusting for mean capture probability and prior standard deviation in a logistic regression model) were lower numbers of passively captured cases, lower number of passively and actively captured cases, and lower number of positive tests among patients tested by active surveillance. The resulting lack of information caused slow mixing but not catastrophic failure. If the threshold for convergence was set to $\hat{R}<1.29$, convergence rates would be 100\%.

The lowest convergence rate of 77\%  was observed for the scenario in which the mean capture probability was 0.75 with the largest prior standard deviation on $\alpha_0$ of 1.5. A high passive surveillance capture probability implies a lower true disease incidence. Since the active surveillance capture probability, $\pi^A$, was held constant across scenarios, the low incidence setting yields far fewer positive tests. Our convergence evaluation found that convergence was strongly associated with the number of positive tests, as these provide more signal for the MCMC chains to mix well.

\begin{figure}[!htb] 
    \centering
    \includegraphics[width=0.8\linewidth]{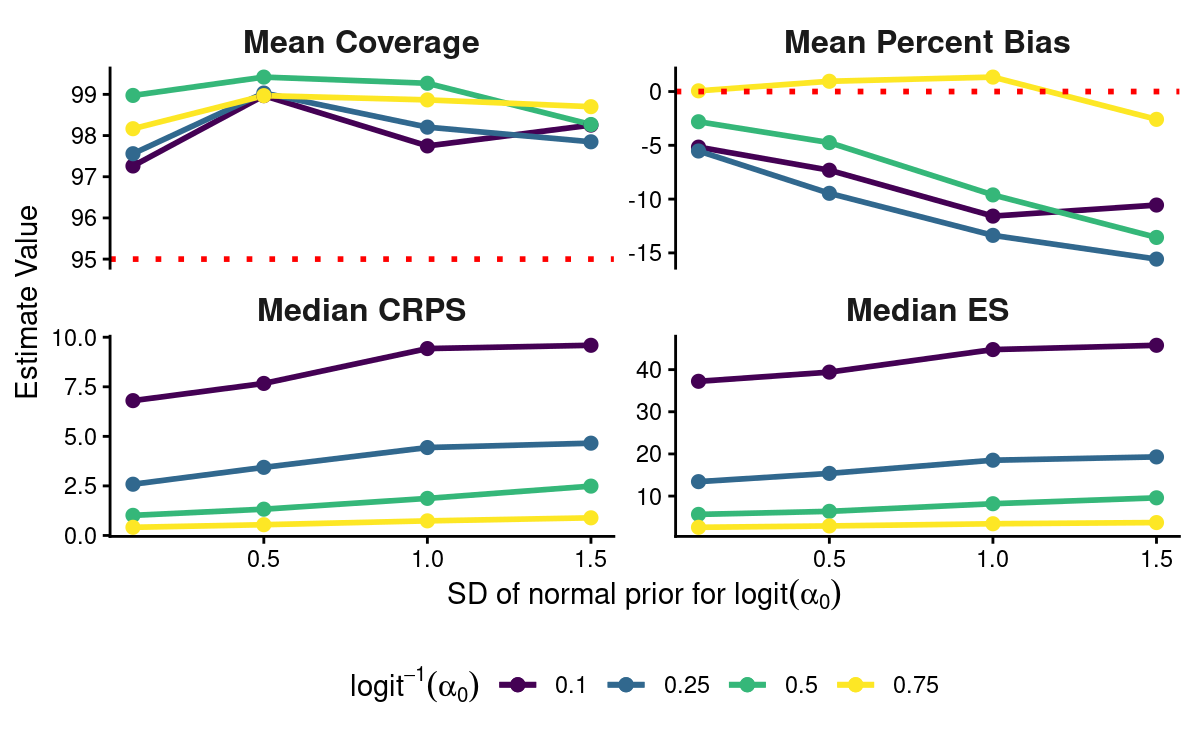}
    \caption{Results of the simulation study with regards to predictive performance. Coverage and the geometric mean percent bias are averaged across all seven regions and simulations. The median CRPS (continuous-ranked probability score) evaluates predictive performance on the island-wide rate of leptospirosis. The median energy score (ES) is the multivariate generalization of the CRPS and considers the predictions for all seven regions simultaneously. $\text{logit}^{-1}(\alpha_0)$ is the baseline capture probability.}
    \label{fig:forecast_results}
\end{figure}

The results of the simulation study with regards to predictive performance are shown in Figure \ref{fig:forecast_results}. We only used the results from simulations for which models converged, excluding those with slow mixing out of an abundance of caution. Predictive $95\%$ credible intervals were generally conservative, with average coverage probability above 95\%. Accounting for under-capture in the proposed model resulted in substantial bias mitigation. For example, with an average case capture probability of $10\%$, we would expect a bias of approximately $-90\%$ in the na\"ive rate estimates since only $10\%$ of cases are captured. However, even with only four hospitals under active surveillance and a relatively uninformative prior on $\alpha_0$, the bias in our proposed model was around $-10\%$. With a strongly informative prior, it was $-5\%$. Bias consistently improved as the prior on $\alpha_0$ became more informative. 

Predictive performance as measured by the CRPS and ES also improved as the informativeness of the prior increased. Additionally, the CRPS and the ES metrics indicated that predictive performance was better for higher values of the mean case capture probability. This is attributable to the sensitivity of the estimated true disease rate to small absolute changes in the estimated case capture probabilities. For example, estimating a case capture probability of 5\% instead of 10\% implies a $20$ times higher true disease rate instead of a $10$ times higher disease rate than observed. Conversely, estimating it to be 70\% instead of 75\% implies a $1.4$ times higher true disease rate instead of a $1.3$ times higher disease rate than observed. Thus, lower capture probabilities result in a more diffuse posterior for the predicted disease rates. Scoring rules like the CRPS and the ES punish this diffuseness. Evaluation of estimation of $\mu_{s,2023}$ and the average probability of capture by the passive surveillance system in each health region were concordant with these results. Plots showing estimates from 100 randomly selected simulations are provided in Section S2.3 of the Supplement.

\begin{figure}[!htb] 
    \centering
    \includegraphics[width=0.8\linewidth]{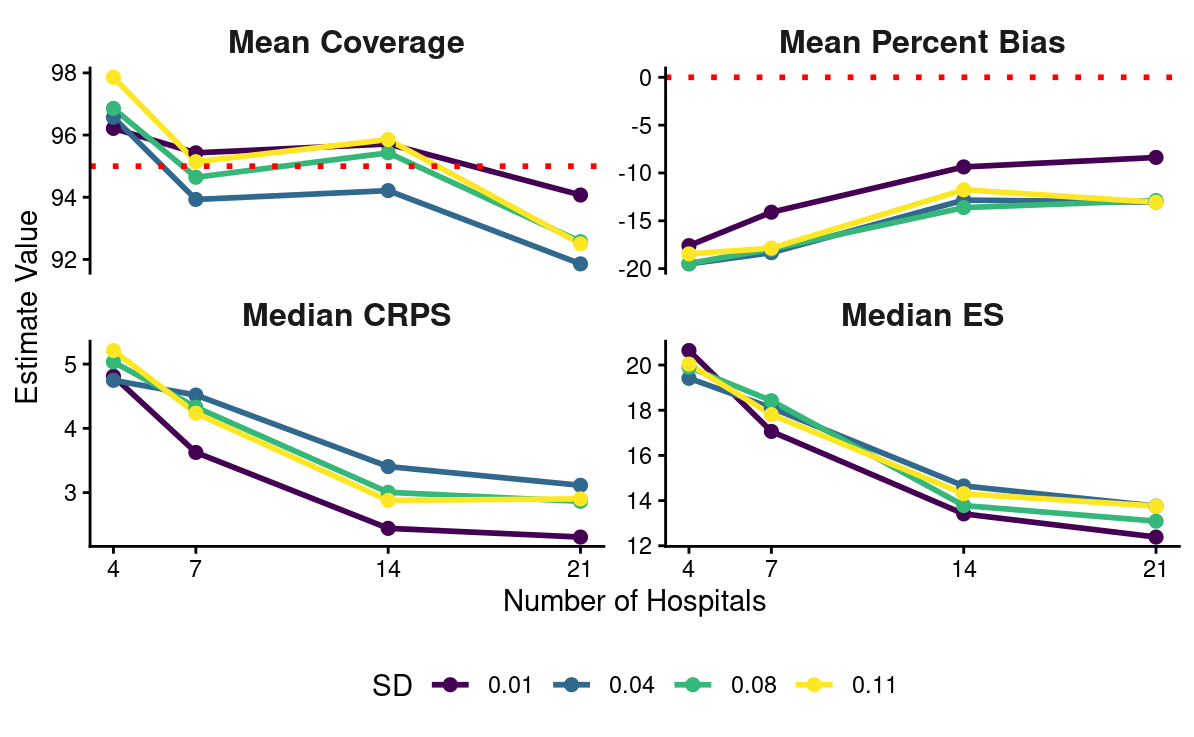}
    \caption{Simulation results with regards to predictive performance for varying numbers of hospitals and priors on test sensitivity and specificity. Coverage and geometric mean percent bias are averaged across all seven regions and simulations. The median CRPS (continuous-ranked probability score) evaluates predictive performance on the island-wide rate of leptospirosis. The ES is the multivariate generalization of the CRPS and considers the predicted rates for all seven regions simultaneously. SD denotes the standard deviation of the beta priors on sensitivity and specificity. An SD of 0.11 was used in the main sensitivity analysis. Lower values of SD correspond to more informative priors.}
    \label{fig:sens}
\end{figure}

In the simulations with increasing numbers of hospitals under active surveillance, convergence rates were above 97\% for all scenarios when seven hospitals were surveilled. When 14 or 21 hospitals were surveilled, convergence rates were virtually 100\%. This provides further evidence that the fragility of convergence is attributable to the paucity of data provided when only actively surveilling four hospitals. The results are shown in Figure \ref{fig:sens}. As we would expect, CRPS and ES improved substantially as the number of surveilled hospitals increased. Bias also generally improved. Coverage declined slightly below nominal levels as the number of hospitals actively surveilled increased. This was likely attributable to the presence of residual bias coupled with tightening credible intervals as more hospitals were surveilled and more data became available. Although the overall trends were not always consistent, the scenario with the strongest priors on sensitivity and specificity generally observed the most appropriate coverage, lowest bias, and lowest CRPS and ES.

\section{Data Application} \label{sec:data}

Returning to the motivating data on leptospirosis, we seek to apply the proposed methodology to correct the observed leptospirosis rates and estimate the true burden of disease in each health region of Puerto Rico. The data is described in Section \ref{sec:leptodata}. Observed rates were as low as 0 per 100,000 in 2022 in Fajardo and as high as 15 per 100,000 in Caguas in 2023. The island-wide rate was 8 per 100,000 in 2022 and 9 per 100,000 in 2023.

\subsection{Model set up}

We adopted a similar model to that used in the simulation study, but with several modifications. First, because information on capacity for a large proportion of the $64$ hospitals is not publicly available, we set $\gamma=2$ and $\delta=0$ when defining the hospital presentation probabilities as described in Equation \ref{eq:gravity}. This specification implies that hospital presentation is a function of motorized travel time only. Furthermore, in the simulation study we allowed the mean function for the Poisson distribution on the true counts to be time varying by incorporating average yearly rainfall. This was impractical for the data analysis since enrollment in the active surveillance study was discontinuous due to disruptions caused by the COVID-19 pandemic \citep{munoz2025diagnosis}. Thus, we adopted the following simplified regression model for the rate of leptospirosis cases presenting to the hospital which was time-invariant:
\begin{align*}
\log(\mu_{s,t}) &= \beta_0 + \Phi_s + \tau_s +  \log(N_s) \\
\Phi &\sim \text{ICAR}(\sigma^2_\Phi), \; \boldsymbol\tau \sim \mathcal{N}(0,\sigma_{\tau}^2I).
\end{align*}

\noindent We used the same model for the hospital-level passive surveillance case capture probabilities as in the simulation study (see Equation \ref{eq:probreg}).

An additional difficulty associated with the discontinuous nature of enrollment for the active surveillance study is specifying the prior for each $n^D_{h,s,t}$, the true number of disease patients choosing each hospital. As discussed earlier, the model implies the following conditional prior distribution for each $n^D_{h,s,t}$:
\begin{align*}
n^D_{h,s,t}|\Theta &\sim \text{Poisson}\left(\mu_{s,t} \times  \pi^C_{h,s}\right)
\end{align*}
\noindent Under the assumption that $\mu_{s,t}$ defines the yearly time-invariant rate for a Poisson process which has independent increments, the distribution of $n^D_{h,s,t}$ in the data application is simply: 
\begin{align*}
n^D_{h,s,t}|\Theta &\sim \text{Poisson}\left(E_h \times \mu_{s,t} \times  \pi^C_{h,s}\right)
\end{align*}
\noindent where $E_h$ is the total enrollment time of hospital $h$ in years.  This adjustment is particularly important in our study, because total enrollment times ranged from $1.5$ months to $18$ months. 

The full list of priors used in the data analysis is provided in Section S3.2 of the Supplement. For $\beta_0$ and the three variance parameters we used weakly informative priors that reflected the following assumptions: 1) the leptospirosis rate is probably much higher than reported but there is still significant uncertainty in the under-reporting rate 2) Heterogeneity in health region-level rates and hospital-level capture probabilities is likely, but, again, there is significant uncertainty. Priors on $\alpha_0$ and $\pi^A$ were more informative. They were guided by prior knowledge \citep{sasaki1993active} and data on enrollment provided by \cite{munoz2025diagnosis}. For test results,  we only considered the IgM rapid test and the PCR test. Informative priors on their sensitivity and specificity were informed by results from the literature \citep{niloofa2015diagnosis, valente2024diagnosis, le2025urine}. This was necessary to ensure the model was identifiable. We ran three chains of 50,000 samples with 10,000 samples discarded as burn-in and used the same sampling strategy as that in the simulation study. Convergence was excellent as assessed by trace plots and the $\hat{R}$ statistic \citep{gelman1992inference}. 

\subsection{Results}

The posterior median for the baseline case capture probability ($\text{logit}^{-1}(\alpha_0)$) of the passive surveillance system was $0.22$ (95\% CI=$0.12$, $0.4$).  Posterior medians of passive surveillance case capture probabilities for individual hospitals in the active surveillance study ranged from $0.21$ to $0.24$ However, the associated credible intervals were very wide. The case capture probability of the active surveillance system ($\pi^{A}$) was estimated to be lower with a posterior median of 0.14 (95\% CI=$0.07$, $0.26$). The model estimated specificity values to be very close to $1$, indicating that false positives were unlikely. This was not the case with sensitivity parameters. Posterior summaries of these and other key parameters are provided in Table S3.2 of the Supplement. Although it's surprising that the estimated active surveillance case capture probability is lower than that of the passive surveillance case capture probability, this result is quite reasonable in our application and is discussed further in the next section.

\begin{figure}[!htb] 
    \centering
    \includegraphics[width=\linewidth]{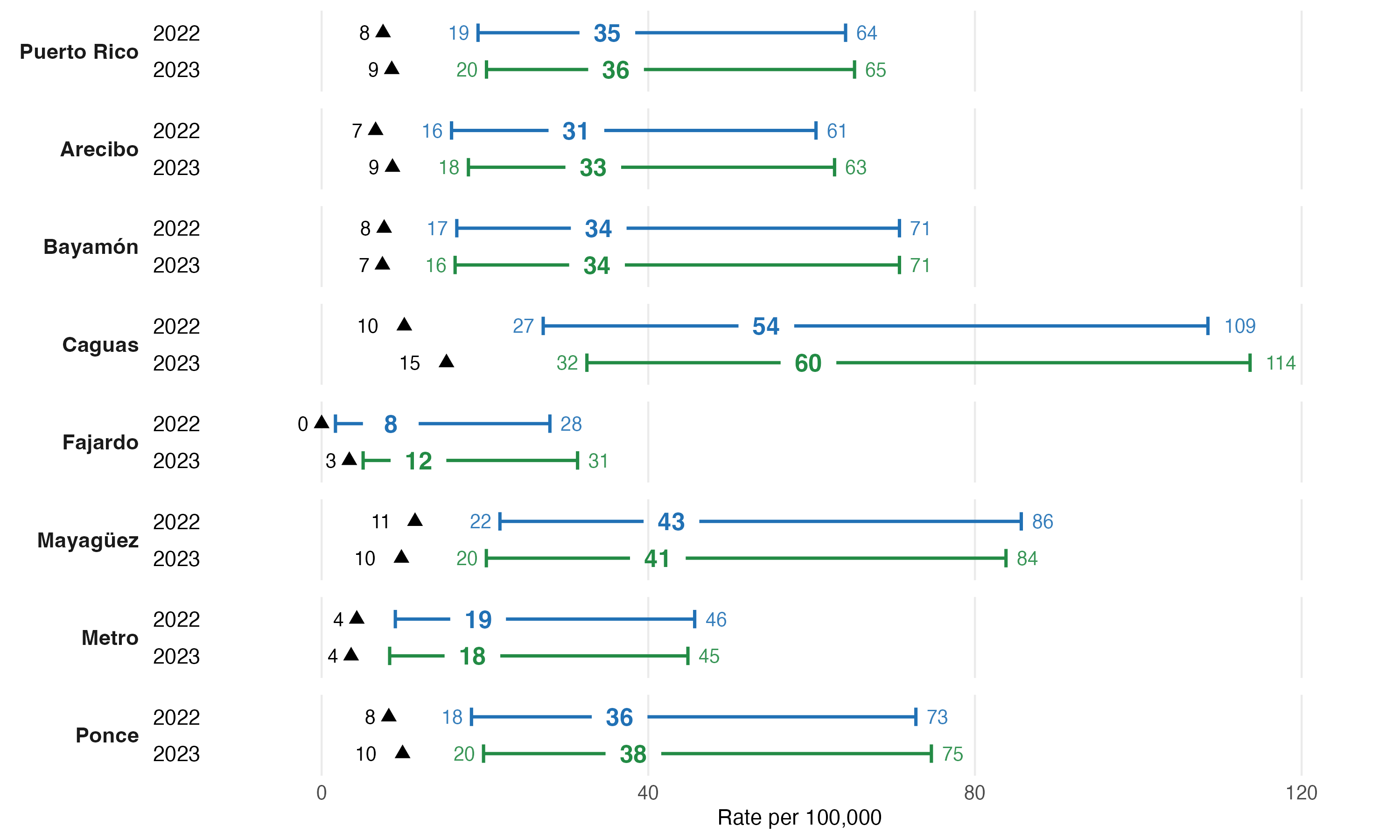}
    \caption{Results of real data analysis: posterior median, lower bound of 95\% prediction interval, and upper bound of 95\% prediction interval for true hospital-presenting leptospirosis rates per 100,000 people in the general population by health region in 2023. Observed rates are denoted by black triangles. Numbers for the lower bound of the prediction intervals are omitted if these values are close to the observed rates}
    \label{fig:2023_results}
\end{figure}

Our primary interest lies in the posterior predictions for the true rate of hospital-presenting leptospirosis in Puerto Rico after adjusting for under-capture by the passive surveillance system. Figure \ref{fig:2023_results} shows the observed rate from passive surveillance and posterior predictive summaries of the estimated rate by health region in 2022 and 2023. The estimated island-wide true rates were 35 per 100,000 (95\% PI=19, 64) in 2022 and 35 per 100,000 (95\% PI=20, 65) in 2023, compared to observed rates of 8 per 100,000 and 9 per 100,000 in those years. True rates were estimated to be as high as 60 per 100,000 in Caguas and as low as 12 per 100,000 in Fajardo in 2023. Despite high variability for most estimates, our model provides strong evidence that taking observed rates at face value results in substantial under-estimation of the true burden of disease. 

\subsection{Sensitivity Analyses}

Examination of the active surveillance test results revealed that there were a relatively large number of patients that tested positive on the IgM rapid test but negative on the PCR test (Table \ref{tab:as_table}). Thus, we anticipate our findings could be sensitive to the informative priors placed on the sensitivity and specificity parameters. In the main analysis, we centered priors on what we felt were the most sensible values based on these past studies. Here, we explore the impact of centering the specificity of the rapid tests on $0.975$, the highest value for an IgM rapid test reported by \cite{niloofa2015diagnosis}, and the sensitivity of the PCR test on $0.7$, the lowest value reported by \cite{le2025urine}. The effect of this is to make the model less skeptical that those patients with a positive IgM rapid test and negative PCR test have the disease. We again ran three chains of 50,000 samples with 10,000 samples discarded as burn-in. Examination of trace plots and $\hat{R}$ statistics indicated the model had converged. For this model, the estimated case capture probability for an average hospital was 0.20 (95\% CI=0.10, 0.36). For the entire island, the estimated leptospirosis rate increased from 35 to 40 per 100,000 in 2022 (95\% CI=20, 82) and 36 to 41 per 100,000 in 2022 (95\% CI=21, 83).

Finally, because sensitivity and specificity parameters are not separately identifiable from the passive surveillance case capture probabilities, we ran a second sensitivity analysis in which the priors for sensitivity and specificity were so strong as to essentially fix these quantities. Just as before, we ran three chains with 50,000 samples with 10,000 samples discarded as burn-in. For the entire island, the estimated rate decreased from 35 to 28 per 100,000 in 2022 (95\% CI=15, 53) and 36 to 27 per 100,000 in 2022 (95\% CI=14, 52). Interestingly, the variability of the estimated passive surveillance capture probabilities noticeably increased with posterior medians ranging from 0.25 to 0.34. Posterior estimates with 95\% credible intervals for both sensitivity analyses are provided in Table S3.3 of the Supplement. 

\section{Discussion} \label{sec:disc}

In this paper, we have presented a novel method for the situation in which active surveillance operates alongside a passive surveillance system in a subset of healthcare facilities, but passive-surveillance captured disease counts are only available at an aggregated level. Rather than considering ideal circumstances, this setting reflects the fact that (1) publicly available surveillance data are unlikely to provide hospital-level case counts and (2) costs and logistics may make it impossible to perform community-based active surveillance. Conducting active surveillance at the hospital-level shifts the our target of estimation to the true rate of hospital-presenting disease cases, rather than the true rate of all disease cases. This is still a meaningful quantification of disease burden and resource utilization. When community-based active surveillance data are available, the methods outlined by \cite{li2020spatial} and \cite{zhang2023hierarchical} can be used. However, incorporation of our methods to account for diagnostic uncertainty would still be helpful.

Importantly, our work has implications for public health decision-making for leptospirosis in Puerto Rico. We acknowledge that the use of informative priors for some parameters means that our results are not entirely data driven. These priors are necessary to ensure identifiability, convergence, and reasonable results given the paucity of available data. A more comprehensive active surveillance study could allow inference to be entirely data driven. Of course, the ability to use prior information where necessary but still allow available data to inform inference is a strength of our Bayesian framework as long as the results are interpreted appropriately. Nevertheless, the weight of the evidence strongly suggests that the rate of hospital-presenting leptospirosis is much higher than reported. Therefore, more effort needs to be invested in surveillance systems and mitigation strategies to better understand the true burden of disease and reduce suffering. For example, public health officials could formulate interventions to increase the rate of leptospirosis testing by clinicians.

One surprising result from our data analysis is that the estimated active surveillance capture probability was lower than the estimated passive surveillance capture probability. However, we believe that our posterior estimates are reasonable. \cite{munoz2025diagnosis} report that 1,689 patients were identified as potentially eligible for the active surveillance project based on emergency department triage logs. Of those, 1,158 patients could be found, 744 were eligible, and 406 were enrolled. Moreover, active surveillance was only conducted 5 days a week. If we assume that eligibility rates are the same among all patients including those not found and that emergency department presentation rates are roughly constant throughout the week, then we estimate that 406 of 1,689 $\times$ 744/1,158 $\times$ 7/5 $\approx$ 1519 UAFI patients were enrolled. 406/1,519 $\approx$ 0.27. Of course, this assumes that probability of inclusion is unrelated to leptospirosis infection. Since patients generally did not know the cause of their symptoms, this is likely reasonable. 

Importantly, this calculation does not account for the fact no surveillance was done from 11 PM to 7 AM, some days only had one shift (either 7 AM to 3 PM or 3 PM to 11 PM), and some eligibility criteria like ability to consent could have excluded true leptospirosis patients. It's also likely that some patients were not identified by project staff since they relied on screening emergency department triage information. Although imprecise, these calculations indicate that our posterior estimate of 0.14 (0.07, 0.26) for $\pi^A$, the active surveillance case capture probability, is reasonable.

Our methodological framework has several key strengths. Most importantly, our model allows for inter-hospital heterogeneity in case capture probabilities. The mechanism used to link estimated hospital-level passive surveillance capture probabilities with the likelihood for hospital-specific and aggregate case counts is rooted in discrete choice methods. Applying these methods leads to choice probabilities resembling a gravity model with concomitant advantages in interpretation. Moreover, we leverage recent work by \cite{weiss_global_2020} to provide more realistic measures of distance incorporating travel time rather than the more typical Euclidean metric. Second, our model allows for imperfect diagnostic test performance, which was not the case in previous work. This feature is especially important in the context of neglected infectious diseases, for which highly-sensitive and specific diagnostic tests may not be available.  Finally, our methods are readily available for public health researchers. We provide posterior predictive distributions that cover different scenarios of interest. These posterior predictive distributions allow for estimation of true hospital-presenting disease counts. The model is also implemented in Nimble, a publicly available software package that is fully contained within the R software ecosystem  \citep{de2017programming}.

One limitation of our approach is that it requires sourcing a data set that encompasses all possible health facilities that a patient could choose. This is necessary to implement the discrete choice methods underlying the gravity model as we must know all choices a patient could make. Although we used a reliable, recently assembled data set of general acute care hospitals \citep{dhs2019hospitals}, it is possible some patients presented to healthcare facilities that are not included in this dataset. This may also present difficulties in the application of our method to areas other than Puerto Rico, especially those with worse infrastructure or limited information about existing infrastructure. Future work could explore the impact of omitting some potential healthcare facilities as well as methods to incorporate uncertainty in $\gamma$ and $\delta$ in the utility model. 

An additional limitation is that we are restricted to a Poisson generating process for the true disease counts. Using the negative binomial distribution to account for potential over-dispersion is not compatible with our framework since binomial/multinomial thinning of negative-binomial distributed random variables does not produce independent random variables. This makes model formulation difficult. However, we note that the inclusion of flexible random effects allowed by our model can often mitigate over-dispersion \citep{stoner2019hierarchical}.

Despite these limitations, our model provides a sensible way to better estimate true disease burden. Even in the presence of sparse data such as in the real data analysis, it can provide useful estimates to better inform public health decision-making with regards to leptospirosis and other neglected pathogens. 

\newpage

\begin{acks}[Acknowledgments]
The authors acknowledge the Minnesota Supercomputing Institute (MSI) at the University of Minnesota for providing resources that contributed to the research results reported within this paper.
\end{acks}

\begin{supplement}
\stitle{Supplement to A Bayesian Spatiotemporal Model to Estimate Disease Burden Using Hospital-Based Active Surveillance}
\sdescription{Supplementary derivations, discrete-choice justification, recursive testing-data likelihood calculation, posterior predictive details, simulation-study details, and additional data-analysis tables and figures.}
\end{supplement}

\begin{supplement}
\stitle{Code and reproducibility files}
\sdescription{R scripts, analysis workflows, and supporting files used to reproduce the analyses in this article. The versioned repository is available at \url{https://github.com/brent-strong/lepto_burden_code}.}
\end{supplement}

\bibliographystyle{imsart-nameyear}
\bibliography{Bib}

\end{document}


\begin{frontmatter}

\title{Supplement to A Bayesian Spatiotemporal Model to Estimate Disease Burden Using Hospital-Based Active Surveillance}
\runtitle{Supplement to Estimating Disease Burden}

\begin{aug}
\author{\fnms{Brent}~\snm{Strong}}
\author{\fnms{Claudia}~\snm{Mu\~noz-Zanzi}}
\author{\fnms{Caitlin}~\snm{Ward}}
\end{aug}

\end{frontmatter}

\section{Supplemental Methods}

\subsection{Derivation of the distribution of $y_{s,t}$ and $z_{s,t} - y_{s,t}$} \label{app:dg_model}

Our model specification is equivalent to assuming the following data generating model:

\begin{align*}
z_{s,t}|\mu_{s,t}, \Theta &\sim \text{Poisson}(N_s \times \mu_{s,t}) \\
\mathbf{z}_{s,t} | z_{s,t}, \Theta &\sim \text{Multinomial}\left(z_{s,t}; \frac{N_{1_s}}{N_s},...,\frac{N_{U_s}}{N_s}\right) \\
\mathbf{n}^D_{u_s,t}|z_{u_s,t},\Theta &\sim \text{Multinomial}\left(z_{u_s,t}; \pi_{h,1_s}^{C},...,\pi_{h,U_s}^{C}\right) \\
n^P_{h,u_s,t} | n^D_{h,u_s,t},\Theta &\sim \text{Binomial}\left(n^D_{h,u_s,t},\pi_{h}^{P}\right)
\end{align*}

\noindent $\mathbf{z}_{s,t}$ (differentiated from $z_{s,t}$ using boldface type) is an $U_s$ dimensional vector where element $u_s$ is the number of cases in the $uth$ subunit of spatial unit $s$. In this model, integrating out $z_{s,t}$, $\mathbf{z}_{s,t}$, and $\mathbf{n}^D_{u_s,t}$ yields the following marginal distribution for $n^P_{u_s,h,t}$:

\begin{align*}
&n^P_{h,u_s,t}| \lambda_{s,t}, \Theta \sim \text{Poisson}\left(\pi_{h}^{P} \times \pi_{h,u_s}^{C} \times \frac{N_{u_s}}{N_s} \times \mu_{s,t}\right) \implies \\
&\sum_{h=1}^H \sum_{u =1}^{U_s} n^P_{h,u_s,t} = \sum_{h=1}^H n^P_{h,s,t} = y_{s,t}  \sim \text{Poisson}\left(\mu_{s,t} \sum_{h=1}^H \pi_{h}^{P} \left[N_s^{-1}\sum_{u=1}^{U_s} N_{u_s} \pi_{h,u_s}^{C}\right] \right) , \; \text{where} \\
&N_s^{-1}\sum_{u=1}^{U_s} N_{u_s} \pi_{h,u_s}^{C} = \pi_{h,s}^C
\end{align*}

\noindent For a proof, see propositions 5.2 and 5.4 of \cite{ross2014introduction}. If we are interested in the unobserved disease patients ($a_{s,t}=z_{s,t}-y_{s,t}$), we consider that: $n^D_{h,u_s,t}-n^P_{h,u_s,t} | n^D_{h,u_s,t},\Theta \sim \text{Binomial}\left(n^D_{h,u_s,t},1-\pi_{h}^{P}\right)$. Summing $n^D_{h,u_s,t}-n^P_{h,u_s,t}$ across all hospitals and spatial subunits in the same manner as before shows that:

\begin{align*}
z_{s,t}-y_{s,t}  \sim \text{Poisson}\left(\mu_{s,t} \sum_{h=1}^H \left(1-\pi_{h}^{P}\right) \left[N_s^{-1}\sum_{u=1}^{U_s} N_{u_s} \pi_{h,u_s}^{C}\right] \right)
\end{align*}

\subsection{Discrete Choice Methods for Hospital Choice Probabilities} \label{app:dc}

In this section, we provide a more formal justification for the form of the hospital choice probabilities, the set of which is denoted by $\{\pi_{h,u_s}^{C}\}_{h=1}^H$. We first work on the individual level. Using the notation of \cite{train_discrete_2009}, we write the utility of person $i$ in spatial subunit $u_s$ if they choose hospital $h$ as:

\begin{align*}
U_{h,i} = V_{h,i} + \epsilon_{h,i}
\end{align*}

\noindent In this formula, $V_{h,i}$ is a deterministic function that is representative of the utility of choosing hospital $h$ (given the need for hospital care) in the population and $\epsilon_{h,i}$ is a stochastic quantity that adjusts individual $i$'s utility from the population average. As outlined in Train (2009), the multinomial probability that person $i$ makes choice $h$ under the assumptions of utility maximization and that the set of $\epsilon_i=(\epsilon_{1,i},...,\epsilon_{H,i})^T$ are independently and identically distributed with a generalized extreme value distribution can be written as:

\begin{align*}
\pi_{h,i}^{C} &= \int\mathbb{I}(\epsilon_{k,i}-\epsilon_{h,i} < V_{h,i} - V_{k,i})\Pr(\epsilon_i)d\epsilon_i \\
&= \int \left( \prod_{k \neq h} e^{-e^{-(\epsilon_{h,i}+V_{h,i} - V_{h,i})}} \right) e^{-\epsilon_{h,i}}e^{-e^{-\epsilon_{h,i}}}
d\epsilon_i \\
&=\frac{e^{V_{h,i}}}{\sum_{k}e^{V_{h,i}}}
\end{align*}

In the case of hospital choice, the following model is sensible:

\begin{align*}
V_{h,i} = G - \gamma \log(\text{distance}_{h,i}) + \delta \log(\text{capacity}_h)
\end{align*}

\noindent where $\text{distance}_{h,i}$ is the distance from where person $i$ is infected to hospital $h$ and $\text{capacity}_h$ is some measure of the capacity of hospital $h$. Since $\text{distance}_{h,i}$ and $\text{capacity}_h$ are assumed to be constant for every individual in spatial subunit $u_s$, we can just consider:

\begin{align*}
\pi_{h,u_s}^{C} = \frac{\text{capacity}_h^{\delta}/\text{distance}_{h,u_s}^{\gamma}}{\sum_{l=1}^{H}\text{capacity}_l^{\delta}/\text{distance}_{l,u_s}^{\gamma}}
\end{align*}

As mentioned in the main text, we fix $\delta$ and $\gamma$. The justification for this is that with only a subset of hospitals under active surveillance at time $t$ (denote this subset by $\mathcal{H}_t$), estimation of $\gamma$ and $\delta$ is difficult because the distribution of $\epsilon_i$ for the patients observed at time $t$ is conditional on $\exists h \in \mathcal{H}_t: \epsilon_{k,i}-\epsilon_{h,i} < V_{h,i} - V_{k,i} \; \forall k \notin \mathcal{H}_t$. Estimation of $\gamma$ and $\delta$ requires linking the now intractable probability model to the observed patient choices.

\subsection{Recursive Algorithm for Testing Data Model} \label{app:rec}

For ease of notation, we drop subscripts  throughout. In the first line of the following equation block, we restate the testing-data likelihood in the main text and then provide its form more explicitly:

\begin{align}
&\Pr(\mathbf{Y}|\Theta,n^{PA},n^{A}) = \sum \Pr(\mathbf{Y}|\boldsymbol{\nu},\boldsymbol{\kappa},\mathbf{d})\Pr(\mathbf{d}|n^{A},n^{PA}) \notag \\
&={n^T-n^{PA} \choose n^{A} - n^{PA}}^{-1} \sum \prod_{j=1}^{n^T}\prod_{k=1}^{K}\left[\nu_k^{y_{jk}d_j}(1-\nu_k)^{(1-y_{jk})d_j}(1-\kappa_k)^{y_{jk}(1-d_j)}\kappa_k^{(1-y_{jk})(1-d_j)}\right]^{\iota_{jk}} \label{eq:explicit}
\end{align}

\noindent where $\iota_{jk}$ is an indicator variable for whether patient $j$ received test $k$. Since we assume that the specificity of the passive surveillance process is $1$, we constrain the first $n^{PA}$ disease status indicators to be $1$. Thus, the sum is across $n^T-n^{PA} \choose n^{A} - n^{PA}$ possible disease status vectors. 

In many cases, calculating the sum will be difficult due to the computational challenges of enumerating all possible combinations of individual disease indicators that sum to $n^{A}$. Fortunately, we can exploit the structural similarity of the sum to the probability mass function of the Poisson binomial distribution to make computation of the exact marginal likelihood feasible. Adopting the notation of \cite{hong2013computing}, the Poisson binomial distribution arises when there are $n$ random indicators distributed as: 
\begin{align*}
I_j \sim \text{Bernoulli}(p_j), \; j=1,...,n
\end{align*}
\noindent and the random variable $N=\sum_{j=1}^nI_j$ is of interest. The probability mass function of $N$ is then: 
\begin{align*}
\Pr(N=k) = \sum_{\mathbf{I} \in \mathcal{I}_k}\prod_{j=1}^np_j^{I_j}(1-p_j)^{1-I_j}
\end{align*}
\noindent where $\mathbf{I}$ is a vector of the random indicators and $\mathcal{I}_k$ is the set of vectors whose elements sum to $k$. Various algorithms have been proposed to exactly compute the probability mass function including discrete Fourier transformation methods and recursive formulae \cite{hong2013computing}. We propose using the recursive method attributed to \cite{barlow} by \cite{hong2013computing}.

Before outlining this approach, we note that Equation \ref{eq:explicit} can be re-written as:
\begin{align*}
&{n^T-n^{PA} \choose n^{A} - n^{PA}}^{-1} \prod_{j=1}^{n^{PA}}\prod_{k=1}^{K}\left[\nu_k^{y_{jk}}(1-\nu_k)^{(1-y_{jk})}\right] \times \\ &\sum \prod_{j=n^{PA}+1}^{n^T}\prod_{k=1}^{K}\left[\nu_k^{y_{jk}d_j}(1-\nu_k)^{(1-y_{jk})d_j}(1-\kappa_k)^{y_{jk}(1-d_j)}\kappa_k^{(1-y_{jk})(1-d_j)}\right]^{\iota_{jk}} \notag 
\end{align*}
\noindent Calculation of the terms proceeding the summation is straightforward. Thus, we only apply the recursive method of \cite{barlow} to the true disease status sub-vectors from index $n^{PA}+1$ to $n^T$. For those patients, we define $w_j^1$ and $w_j^0$ as follows:
\begin{align*}
w_j^1&=\prod_{k=1}^K\left[(\nu_k)^{y_{n^{PA}+j,k}}(1-\nu_k)^{(1-y_{n^{PA}+j,k})}\right]^{\iota_{n^{PA}+j,k}} \\
w_j^0&=\prod_{k=1}^K\left[(1-\kappa_k)^{y_{n^{PA}+j,k}}\kappa_k^{(1-y_{n^{PA}+j,k})}\right]^{\iota_{n^{PA}+j,k}}
\end{align*}
\noindent Conditional on values of sensitivity and specificity (either as fixed values or the current values in the Markov chain), all quantities are known. Let:
\begin{align*}
\xi_{m,l}= \sum_{\mathbf{z}\in \mathcal{D}_{m,l}} \prod_{j=n^{PA}+1}^{n^{PA}+l}\prod_{k=1}^{K}\left[\nu_k^{y_{jk}d_j}(1-\nu_k)^{(1-y_{jk})d_j}(1-\kappa_k)^{y_{jk}(1-d_j)}\kappa_k^{(1-y_{jk})(1-d_j)}\right]^{\iota_{jk}}
\end{align*}
\noindent where $\mathcal{D}_{m,l}$ is the set of sub-vectors corresponding to indices $n^{PA}+1$ to $n^{PA}+l$ with exactly $m$ patients with a disease status of $1$. If we set $\xi_{-1,l}=\xi_{l+1,l}=0$ and $\xi_{0,0}=1$ then the following recursion can be used to compute the desired summation:
\begin{align*}
\xi_{m,l}=w_j^{0}\xi_{m,l-1} + w_j^{1}\xi_{m-1,l-1}
\end{align*}
The complexity of the algorithm is $O(n^2)$, making it feasible to calculate the likelihood exactly for realistic values of $n^T$. If sensitivity and specificity are fixed, the probability distribution for each bin does not need to be re-computed at each iteration of the Markov chain. However, it will generally be cheap to do so.

\subsubsection{Simple Example Calculation}

As a simple example, consider a scenario where $n^T-n^{PA}=2$ and therefore $\xi_{0,2}$, $\xi_{1,2}$, and $\xi_{2,2}$ are of interest. To reiterate, $w_j^1$ and $w_j^0$ are defined as:

\begin{align*}
w_j^1&=\left[\prod_{k=1}^K(\nu_k)^{y_{n^{RT+j},k}}(1-\nu_k)^{(1-y_{n^{PA}+j,k})}\right]^{\iota_{n^{PA}+j,k}} \\
w_j^0&=\prod_{k=1}^K\left[(1-\kappa_k)^{y_{n^{PA}+j,k}}\kappa_k^{(1-y_{n^{PA}+j,k})}\right]^{\iota_{n^{PA}+j,k}}
\end{align*}

\noindent and the recursion is defined as:

\begin{align*}
\xi_{m,l}=w_j^{0}\xi_{m,l-1} + w_j^{1}\xi_{m-1,l-1}
\end{align*}

We show only the calculation of $\xi_{1,2}$. First:

\begin{align*}
\xi_{0,1} &= w_1^{0}\xi_{0,0}+ w_1^{1}\xi_{-1,0}=\prod_{k=1}^K\left[(1-\kappa_k)^{y_{n^{PA}+1,k}}\kappa_k^{(1-y_{n^{PA}+1,k})}\right]^{\iota_{n^{PA}+1,k}} \\
\xi_{1,1} &= w_1^{0}\xi_{1,0}+ w_1^{1}\xi_{0,0}=\prod_{k=1}^K\left[\nu_k^{y_{n^{PA}+1,k}}(1-\nu_k)^{(1-y_{n^{PA}+1,k})}\right]^{\iota_{n^{PA}+1,k}}
\end{align*}

\noindent Therefore:

\begin{align*}
\xi_{1,2} &= w_2^{0}\xi_{1,1}+ w_2^{1}\xi_{0,1} \\
&=\prod_{k=1}^K\left[(1-\kappa_k)^{y_{n^{PA}+2,k}}\kappa_k^{(1-y_{n^{PA}+2,k})}\right]^{\iota_{n^{PA}+2,k}}\prod_{k=1}^K\left[\nu_k^{y_{n^{PA}+1,k}}(1-\nu_k)^{(1-y_{n^{PA}+1,k})}\right]^{\iota_{n^{PA}+1,k}} \\
&+ \prod_{k=1}^K\left[(\nu_k)^{y_{n^{PA}+2,k}}(1-\nu_k)^{(1-y_{n^{PA}+2,k})}\right]^{\iota_{n^{PA}+2,k}}\prod_{k=1}^K\left[(1-\kappa_k)^{y_{n^{PA}+1,k}}\kappa_k^{(1-y_{n^{PA}+1,k})}\right]^{\iota_{n^{PA}+1,k}}
\end{align*}

\noindent This sum is intuitively correct as the first term corresponds to when the disease indicator is $1$ in the second position and the second term corresponds to when the disease indicator is $1$ in the first position, which are the only two possibilities.

\subsection{Posterior Predictive Distributions} \label{app:ppd}

Here, we discuss a scenario in which we are interested in true disease counts during time periods in which we have conducted active surveillance at a subset of hospitals. In this scenario, we should not marginalize out each $n^{D}_{h,s,t}$ so we can use posterior draws of this parameter as part of estimation. It's helpful to write $z_{s,t}$ as follows:

\begin{align}
z_{s,t}= \sum_{h \notin \mathcal{H}^A_t}n^P_{h,s,t} + \sum_{h \notin \mathcal{H}^A_t}n^{D}_{h,s,t}-n^P_{h,s,t} + \sum_{h \in \mathcal{H}^A_t}n^{D}_{h,s,t} \label{eq:post_pred_decomp}
\end{align}

If we have access to aggregate case counts from the same time period as the active surveillance data, then $\sum_{h \notin \mathcal{H}^A_t}n^P_{h,s,t}$ is observed. For each draw from the posterior distribution on parameters, we can draw $\sum_{h \notin \mathcal{H}^A_t}n^{D}_{h,s,t}-n^P_{h,s,t}$ as follows:

\begin{align*}
\sum_{h \notin \mathcal{H}^A_t}n^{D}_{h,s,t}-n^P_{h,s,t} \sim \text{Poisson}\left(\mu^{(b)}_{s,t} \times  \sum_{h \notin \mathcal{H}_t}\pi^C_{h,s}\left(1-{\pi}_{h}^{P}\right)\right)
\end{align*}

\noindent If aggregate case counts are not available then the entirety of $\sum_{h \notin \mathcal{H}^A_t}n^P_{h,s,t} + \sum_{h \notin \mathcal{H}^A_t}n^{D}_{h,s,t}-n^P_{h,s,t} = \sum_{h \notin \mathcal{H}^A_t}n^D_{h,s,t}$ is unobserved. We can draw $\sum_{h \notin \mathcal{H}^A_t}n^D_{h,s,t}$ as follows:

\begin{align*}
\sum_{h \notin \mathcal{H}^A_t}n^{D}_{h,s,t} \sim \text{Poisson}\left(\mu^{(b)}_{s,t} \times  \sum_{h \notin \mathcal{H}_t}\pi^C_{h,s}\right)
\end{align*}

\noindent Using the appropriate combination of samples and observed quantities we can easily calculate posterior predictive draws for $z_{s,t}$ using Equation \ref{eq:post_pred_decomp}. 

\noindent 

\clearpage

\section{Simulation Study}

\subsection{Simulation Study Evaluation Metrics} \label{sec:sim_metrics}

Mean percent bias was calculated using the following formula:
\begin{align*}
\text{Mean Percent Bias} = 100 \times \left(\exp\left\{\frac{1}{7M}\sum_{m=1}^M\sum_{s=1}^7\log\left(\frac{\bar{\tilde{z}}_{s,2024}^{m}/N_s}{z_{s,2024}/N_s}\right)\right\}-1\right)
\end{align*}
\noindent where $m$ indexes simulations and $\bar{\tilde{z}}_{s,2024}^{m}/N_s$ is the median posterior prediction for the true rate in spatial unit $s$ in 2024. $z_{s,2024}/N_s$ is the true rate itself. 

Another metric used to measure forecast performance was the energy score (ES). We denote the concatenated vector of populations in each spatial unit measured in units of 100,000 people as $\mathbf{N}$. The ES takes the following form where $\frac{\mathbf{z}_{2024}}{\mathbf{N}}$ is the true vector of disease rates, $P_{\mathbf{z}_{2024}/\mathbf{N}}$ is the predictive distribution, and $\frac{\widetilde{\mathbf{z}}_{2024}}{\mathbf{N}}$ and $\frac{\widetilde{\mathbf{z}}'_{2024}}{\mathbf{N}}$ are independent draws from this predictive distribution:
\begin{align}
ES\left(P_{\mathbf{z}_{2024}/\mathbf{N}},\frac{\mathbf{z}_{2024}}{\mathbf{N}}\right)= \mathbb{E}\left[\left\|\frac{\widetilde{\mathbf{z}}_{2024}}{\mathbf{N}}-\frac{\mathbf{z}_{2024}}{\mathbf{N}}\right\|\right]-\frac{1}{2}\mathbb{E}\left[\left\|\frac{\widetilde{\mathbf{z}}_{2024}}{\mathbf{N}}-\frac{\widetilde{\mathbf{z}}'_{2024}}{\mathbf{N}}\right\|\right] \label{eq:ES}
\end{align}
\noindent For simulation $m$, we can estimate $ES(P_{\mathbf{z}_{2024}/\mathbf{N}},\frac{\mathbf{z}_{2024}}{\mathbf{N}})$ using the $L$ draws from the posterior predictive distribution and the following estimator: 
\begin{align*}
\widehat{ES}_m\left(P_{\mathbf{z}_{2024}/\mathbf{N}},\frac{\mathbf{z}_{2024}}{\mathbf{N}}\right)=\frac{1}{L}\sum_{l=1}^L\left\|\frac{\widetilde{\mathbf{z}}^{(l)}_{2024}}{\mathbf{N}}-\frac{\mathbf{z}_{2024}}{\mathbf{N}}\right\|-\frac{1}{L(L-1)}\sum_{1\leq l<l' \leq L}\left\|\frac{\widetilde{\mathbf{z}}^{(l)}_{2024}}{\mathbf{N}}-\frac{\widetilde{\mathbf{z}}^{(l')}_{2024}}{\mathbf{N}}\right\|
\end{align*}
\noindent We can then calculate $\widehat{ES}(P_{\mathbf{z}_{2024}},\mathbf{z}_{2024})$ by taking the median across the $M$ simulations for each scenario to compare model performance. We also used the continuous-ranked probability score (CRPS) to measure the quality of island wide rate predictions. The true island wide rate is defined as follows:  

\begin{align*}
\frac{\sum_{s=1}^Sz_{s,2024}}{\sum_{s=1}^SN_s}
\end{align*}

\noindent The CRPS is a univariate measure that simply involves replacing the distance measure in Equation \ref{eq:ES} with the absolute value of the difference \citep{gneiting2007strictly}.

\subsection{Priors for Simulation Study} \label{sec:sim}

For the simulation study, we used the following priors:

\begin{align*}
\beta_0 &\sim N(3.3,2^2) \\
\beta_1 &\sim N(0,1) \\
\log(\sigma_{\Phi}) &\sim N(0,1) \\
\log(\sigma_{\tau}) &\sim N(0,1) \\
\alpha_0 &\sim N(\mu_{\alpha_0},\sigma_{\alpha_0}^2) \\
\log(\sigma_{\phi}) &\sim N(0,1) \\
\zeta &= 0.06 \\
\pi^{A} &\sim \text{Beta}(5,15) \\
\nu_1 &\sim \text{Beta}(8.5,1.5) \\
\nu_2 &\sim \text{Beta}(8,2) \\
\kappa_1 &\sim \text{Beta}(8.5,1.5) \\
\kappa_2 &\sim \text{Beta}(9.5,0.5) \\
\end{align*}
\clearpage
\subsection{Additional Simulation Study Results} 

\begin{center}
\refstepcounter{figure}
\label{fig:plot_low_prob.png}
\includegraphics[width=\linewidth]{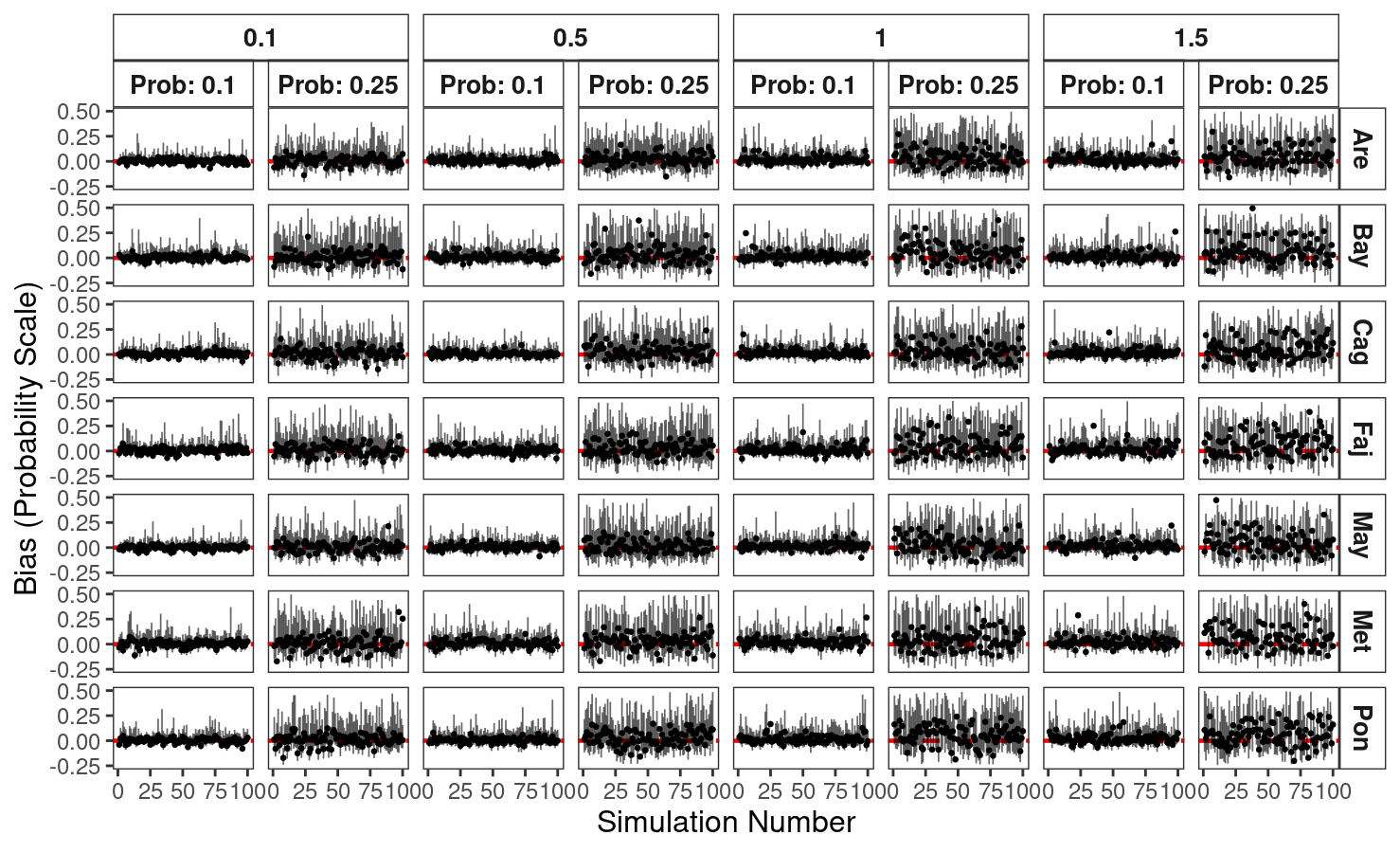}
\smallskip
\begin{minipage}{0.92\linewidth}
\footnotesize {\scshape Fig.}~\thefigure. Results for 100 randomly selected individual simulations from the main simulation study for $\text{logit}^{-1}(\alpha_0)=0.1$ and $\text{logit}^{-1}(\alpha_0)=0.25$. Posterior medians with 95\% credible intervals are shown for the average probability of case capture. The true value has been subtracted so that the results are on the bias scale (a value of 0 corresponds to perfect estimation). The informativeness of the prior on $\alpha_0$ increases from left to right.
\end{minipage}
\end{center}
\clearpage

\begin{center}
\refstepcounter{figure}
\label{fig:plot_high_prob.png}
\includegraphics[width=\linewidth]{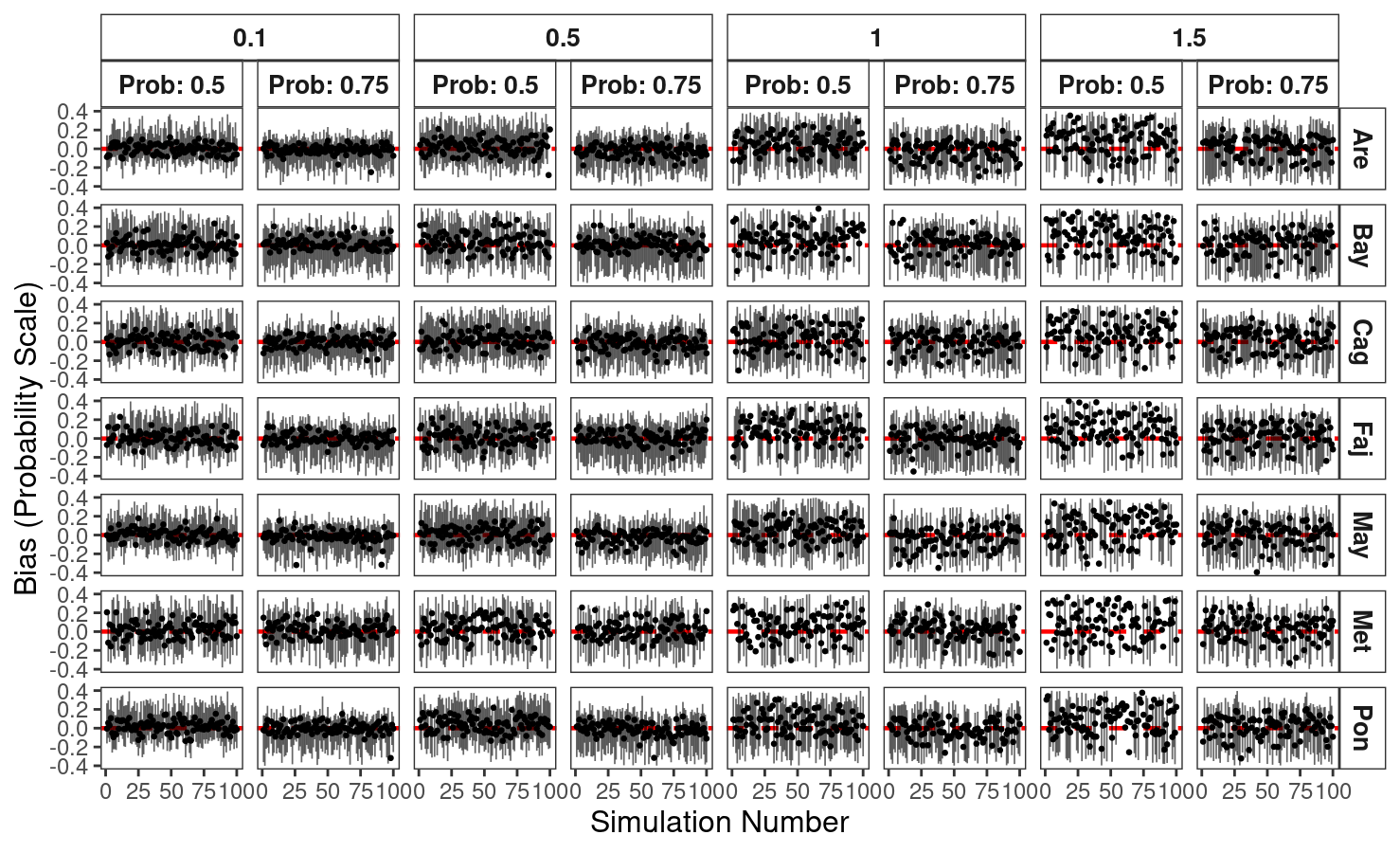}
\smallskip
\begin{minipage}{0.92\linewidth}
\footnotesize {\scshape Fig.}~\thefigure. Results for 100 randomly selected individual simulations from the main simulation study for $\text{logit}^{-1}(\alpha_0)=0.5$ and $\text{logit}^{-1}(\alpha_0)=0.75$. Posterior medians with 95\% credible intervals are shown for the average probability of case capture for each health region. The true value has been subtracted so that the results are on the bias scale (a value of 0 corresponds to perfect estimation). The informativeness of the prior on $\alpha_0$ increases from left to right.
\end{minipage}
\end{center}
\clearpage

\begin{center}
\refstepcounter{figure}
\label{fig:plot_low_rate.png}
\includegraphics[width=\linewidth]{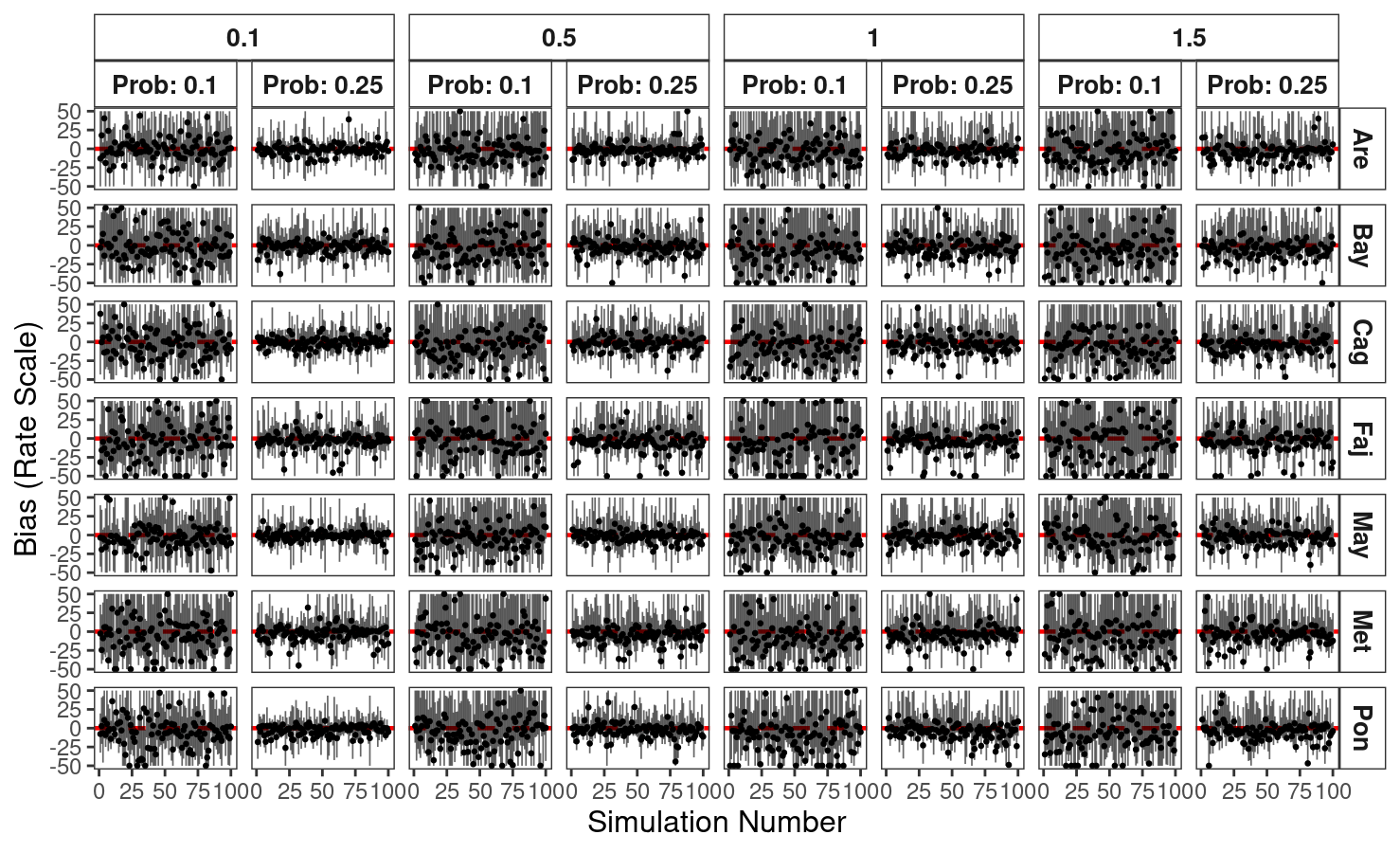}
\smallskip
\begin{minipage}{0.92\linewidth}
\footnotesize {\scshape Fig.}~\thefigure. Results for 100 randomly selected individual simulations from the main simulation study for $\text{logit}^{-1}(\alpha_0)=0.1$ and $\text{logit}^{-1}(\alpha_0)=0.25$. Posterior medians with 95\% credible intervals are shown for the mean rate parameter in 2023 for each health region. The true value has been subtracted so that the results are on the bias scale (a value of 0 corresponds to perfect estimation). The informativeness of the prior on $\alpha_0$ increases from left to right.
\end{minipage}
\end{center}
\clearpage

\begin{center}
\refstepcounter{figure}
\label{fig:plot_high_rate.png}
\includegraphics[width=\linewidth]{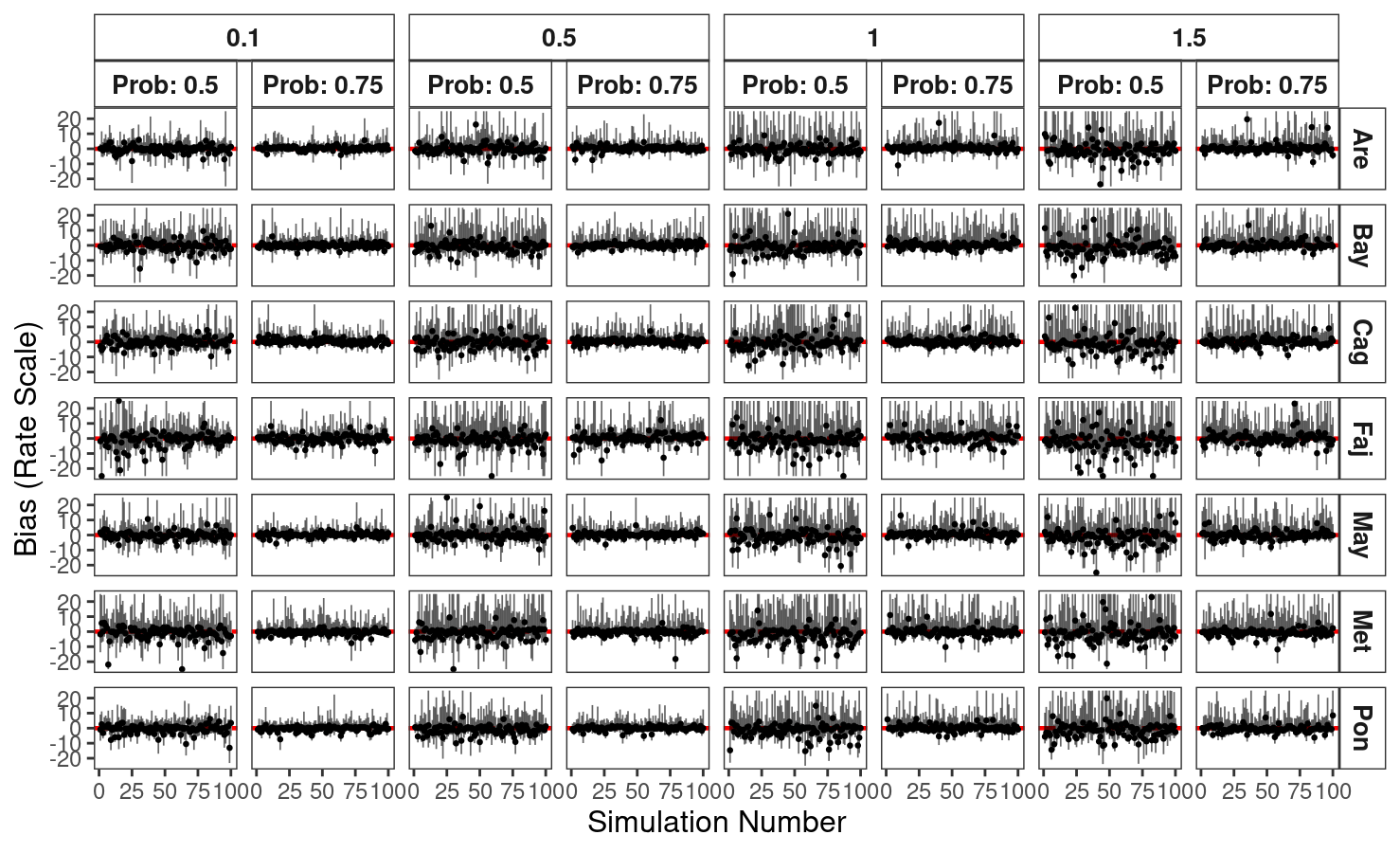}
\smallskip
\begin{minipage}{0.92\linewidth}
\footnotesize {\scshape Fig.}~\thefigure. Results for 100 randomly selected individual simulations from the main simulation study for $\text{logit}^{-1}(\alpha_0)=0.5$ and $\text{logit}^{-1}(\alpha_0)=0.75$. Posterior medians with 95\% credible intervals are shown for the mean rate parameter in 2023 for each health region. The true value has been subtracted so that the results are on the bias scale (a value of 0 corresponds to perfect estimation). The informativeness of the prior on $\alpha_0$ increases from left to right.
\end{minipage}
\end{center}
\clearpage

\section{Data Analysis}

\subsection{Aggregate Case Count Data from 2022 and 2023}

\begin{center}
\refstepcounter{table}
\label{tab:cases_rates}
\begin{minipage}{\linewidth}
\centering
\footnotesize {\scshape Table}~\thetable. The observed rate of leptospirosis (from the passive surveillance system) for each of the seven health regions of Puerto Rico in 2022 and 2023. Case numbers are shown in parentheses.\par\medskip
\begin{tabular}{lccccccc}
\hline
Year & Arecibo & Bayam\'on & Caguas & Fajardo & Mayag\"uez & Metro & Ponce \\
\hline
2022 & 7 (32) & 8 (37) & 10 (53) & 0 (0) & 11 (55) & 4 (31) & 8 (39) \\
2023 & 9 (42) & 7 (36) & 15 (80) & 3 (4) & 10 (47) & 4 (26) & 10 (47) \\
\hline
\end{tabular}
\end{minipage}
\end{center}
\clearpage

\subsection{Prior and Posterior Distributions} \label{sec:pp_dist}

For the data analysis, we used the following priors:

\begin{align*}
\beta_0 &\sim N(3.7,2^2) \\
\log(\sigma_{\Phi}) &\sim N(0,1) \\
\log(\sigma_{\tau}) &\sim N(0,1) \\
\alpha_0 &\sim N(-1.4,0.5^2) \\
\log(\sigma_{\phi_h}) &\sim N(0,1) \\
\zeta &= 0.06 \\
\pi^{A} &\sim \text{Beta}(5,20) \\
\nu_1 &\sim \text{Beta}(8.5,1.5) \\
\nu_2 &\sim \text{Beta}(8,2) \\
\kappa_1 &\sim \text{Beta}(8.5,1.5) \\
\kappa_2 &\sim \text{Beta}(9.5,0.5) \\
\end{align*}

The table below provides posterior summaries for the main analysis. The table on the next page provides results from the two sensitivity analyses.

\begin{center}
\refstepcounter{table}
\label{tab:posterior_results}
\begin{minipage}{\linewidth}
\centering
\footnotesize {\scshape Table}~\thetable. Posterior Medians and 95\% Credible Intervals on Original and Transformed Scales\par\medskip
\renewcommand{\arraystretch}{1.15}
\begin{tabular}{lll}
\toprule
\textbf{Parameter} & \textbf{Original Scale (Median [95\% CI])} & \textbf{Transformed Scale (Median [95\% CI])} \\ \midrule
$\beta_0$ (log)              & 3.40 [2.63, 4.16]    & 30 [14, 64] \\
$\sigma_{\Phi}$ (log)        & -0.43 [-2.09, 0.62]  & 0.65 [0.12, 1.86] \\
$\sigma_{\tau}$ (log)        & -0.75 [-2.08, 0.20]  & 0.47 [0.13, 1.22] \\
$\alpha_0$ (logit)           & -1.26 [-2.01, -0.40] & 0.22 [0.12, 0.40] \\
$\sigma_{\phi}$ (log)        & -0.81 [-2.36, 0.51]  & 0.45 [0.09, 1.67] \\
$\pi^A$ (probability)        & 0.14 [0.07, 0.26]    & 0.14 [0.07, 0.26] \\
$\pi^P_{\text{Bella Vista}}$ (logit) & -1.17 [-2.54, 0.41] & 0.24 [0.07, 0.60] \\
$\pi^P_{\text{Castaner}}$ (logit)    & -1.29 [-2.48, -0.06] & 0.22 [0.08, 0.48] \\
$\pi^P_{\text{DCH}}$ (logit)         & -1.34 [-2.53, -0.14] & 0.21 [0.07, 0.46] \\
$\pi^P_{\text{Ryder}}$ (logit)       & -1.15 [-2.10, 0.00]  & 0.24 [0.11, 0.50] \\
$\nu_1$ (sensitivity IgM)    & 0.87 [0.60, 0.99]    & 0.87 [0.60, 0.99] \\
$\nu_2$ (sensitivity PCR)    & 0.76 [0.49, 0.94]    & 0.76 [0.49, 0.94] \\
$\kappa_1$ (specificity IgM) & 0.98 [0.97, 1.00]    & 0.98 [0.97, 1.00] \\
$\kappa_2$ (specificity PCR) & 0.99 [0.98, 1.00]    & 0.99 [0.98, 1.00] \\ \bottomrule
\end{tabular}
\renewcommand{\arraystretch}{1}
\end{minipage}
\end{center}
\clearpage

\begin{center}
\refstepcounter{table}
\label{tab:posterior_results_sensitivity}
\begin{minipage}{\linewidth}
\centering
\footnotesize {\scshape Table}~\thetable. Posterior Medians and 95\% Credible Intervals for Two Sensitivity Analyses\par\medskip
\renewcommand{\arraystretch}{1.15}
\begin{tabular}{lll}
\toprule
\textbf{Parameter} & \textbf{Original Scale (Median [95\% CI])} & \textbf{Transformed Scale (Median [95\% CI])} \\
\midrule

\multicolumn{3}{l}{\textit{First sensitivity analysis}} \\
$\beta_0$ (log)              & 3.54 [2.81, 4.30]    & 34 [17, 74] \\
$\sigma_{\Phi}$ (log)        & -0.40 [-1.97, 0.66]  & 0.67 [0.14, 1.93] \\
$\sigma_{\tau}$ (log)        & -0.81 [-2.26, 0.17]  & 0.44 [0.10, 1.19] \\
$\alpha_0$ (logit)           & -1.40 [-2.18, -0.57] & 0.20 [0.10, 0.36] \\
$\sigma_{\phi}$ (log)        & -0.85 [-2.44, 0.47]  & 0.43 [0.09, 1.60] \\
$\pi^A$ (probability)        & 0.15 [0.07, 0.28]    & 0.15 [0.07, 0.28] \\
$\pi^P_{\text{Bella Vista}}$ (logit) & -1.36 [-2.35, -0.21] & 0.20 [0.09, 0.45] \\
$\pi^P_{\text{Castaner}}$ (logit)    & -1.43 [-2.64, -0.18] & 0.19 [0.07, 0.46] \\
$\pi^P_{\text{DCH}}$ (logit)         & -1.47 [-2.66, -0.24] & 0.19 [0.07, 0.44] \\
$\pi^P_{\text{Ryder}}$ (logit)       & -1.29 [-2.60, 0.32]  & 0.22 [0.07, 0.58] \\
$\nu_1$ (sensitivity IgM)    & 0.88 [0.63, 0.99]    & 0.88 [0.63, 0.99] \\
$\nu_2$ (sensitivity PCR)    & 0.62 [0.36, 0.86]    & 0.62 [0.36, 0.86] \\
$\kappa_1$ (specificity IgM) & 0.99 [0.97, 1.00]    & 0.99 [0.97, 1.00] \\
$\kappa_2$ (specificity PCR) & 0.99 [0.98, 1.00]    & 0.99 [0.98, 1.00] \\

\midrule
\multicolumn{3}{l}{\textit{Second sensitivity analysis}} \\
$\beta_0$ (log)              & 3.14 [2.36, 3.91]    & 23 [11, 50] \\
$\sigma_{\Phi}$ (log)        & -0.41 [-2.05, 0.72]  & 0.67 [0.13, 2.05] \\
$\sigma_{\tau}$ (log)        & -0.84 [-2.27, 0.19]  & 0.43 [0.10, 1.21] \\
$\alpha_0$ (logit)           & -1.00 [-1.85, -0.13] & 0.27 [0.14, 0.47] \\
$\sigma_{\phi}$ (log)        & -0.61 [-2.29, 0.93]  & 0.54 [0.10, 2.52] \\
$\pi^A$ (probability)        & 0.11 [0.05, 0.23]    & 0.11 [0.05, 0.23] \\
$\pi^P_{\text{Bella Vista}}$ (logit) & -0.66 [-1.77, 1.66] & 0.34 [0.15, 0.84] \\
$\pi^P_{\text{Castaner}}$ (logit)    & -1.02 [-2.37, 0.60] & 0.27 [0.09, 0.65] \\
$\pi^P_{\text{DCH}}$ (logit)         & -1.08 [-2.42, 0.48] & 0.25 [0.08, 0.62] \\
$\pi^P_{\text{Ryder}}$ (logit)       & -0.87 [-2.53, 1.41] & 0.29 [0.07, 0.80] \\
$\nu_1$ (sensitivity IgM)    & 0.85 [0.84, 0.86]    & 0.85 [0.84, 0.86] \\
$\nu_2$ (sensitivity PCR)    & 0.80 [0.79, 0.81]    & 0.80 [0.79, 0.81] \\
$\kappa_1$ (specificity IgM) & 0.85 [0.85, 0.86]    & 0.85 [0.85, 0.86] \\
$\kappa_2$ (specificity PCR) & 0.95 [0.95, 0.96]    & 0.95 [0.95, 0.96] \\

\bottomrule
\end{tabular}
\renewcommand{\arraystretch}{1}
\end{minipage}
\end{center}
\clearpage

\bibliographystyle{imsart-nameyear}
\bibliography{Bib}